%% file: main.tex
\documentclass[journal]{IEEEtran}
\usepackage{titlesec}
\titlespacing*{\section}{0pt}{*1}{*1}
\titlespacing*{\subsection}{0pt}{*0.5}{*0.5}
\usepackage{multirow}
\usepackage[hidelinks,colorlinks=true,linkcolor=blue,citecolor=blue,urlcolor=black]{hyperref}
\usepackage{mathtools}
\usepackage{amsmath,amssymb,amsfonts}
\usepackage{graphicx}
\usepackage{lettrine}
\usepackage{textcomp}
\usepackage{xcolor}
\usepackage{subfigure} 
\usepackage[letterpaper, left=0.65in, right=0.65in, bottom=1in, top=0.70in]{geometry}

\def\BibTeX{{\rm B\kern-.05em{\sc i\kern-.025em b}\kern-.08em
    T\kern-.1667em\lower.7ex\hbox{E}\kern-.125emX}}
\usepackage{xcolor}
\usepackage[linesnumbered,ruled,vlined]{algorithm2e}
\usepackage{listings}
\usepackage[noend]{algpseudocode}

\SetCommentSty{mycommfont}
\SetKwInput{KwInput}{Input}   
\SetKwInput{KwOutput}{Output} 

\definecolor{GreenForest}{rgb}{0.09, 0.45, 0.27}
\usepackage{soul}
\usepackage{balance}
\usepackage{cite}
\usepackage{acro} 
\input{acro.tex}
\usepackage{comment}
\usepackage{amsmath}
\usepackage{steinmetz}
\makeatletter
\newcommand{\ie}{\emph{i.e.}\@ifnextchar.{\!\@gobble}{}}
\newcommand{\eg}{\emph{e.g.}\@ifnextchar.{\!\@gobble}{}}
\newcommand{\etc}{etc\@ifnextchar.{}{.\@}}
\makeatother
\begin{document}

    \title{Redefining Orthogonal Co-Existence: A Mother Waveform Framework for DFT-Based Waveforms}
  \author{Ayoub Ammar Boudjelal, Rania Yasmine Bir, and H\"{u}seyin Arslan,~\IEEEmembership{Fellow,~IEEE}

\thanks{The authors are with the Department of Electrical and Electronics Engineering, Istanbul Medipol University, Istanbul, 34810, Turkey (e-mail: 
ayoub.ammar@std.medipol.edu.tr; rania.bir@std.medipol.edu.tr;    huseyinarslan@medipol.edu.tr).}
}

\IEEEpeerreviewmaketitle
\maketitle

\begin{abstract}
In this paper, we introduce the concept of a "mother waveform" to address key challenges in 5th generation (5G) and 6th generation (6G) networks, including spectrum efficiency, backward compatibility, enhanced flexibility, and the integration of joint sensing and communication (JSAC). We propose single-
carrier interleaved frequency division multiplexing (SC-IFDM) as the mother waveform and demonstrate, through rigorous mathematical modeling, that it can generate all discrete Fourier
transform (DFT)-based waveforms without requiring structural modifications. Specifically, by selectively configuring lattice indices and phase adjustments, SC-IFDM enables seamless adaptation to diverse waveforms, such as orthogonal frequency division
multiplexing (OFDM), orthogonal chirp division multiplexing (OCDM), orthogonal time-frequency space (OTFS), affine frequency division multiplexing (AFDM), and frequency
modulated continuous wave (FMCW) within a unified framework. Critical aspects such as coexistence strategies, and resource allocation   are thoroughly explored. Simulation results demonstrate the proposed framework’s ability to deliver superior communication performance, robust sensing capabilities, and efficient coexistence, surpassing traditional waveform designs in scalability and adaptability.
\end{abstract}


\section{Introduction}

\IEEEPARstart{A}{s} wireless communication progresses toward \ac{6G}, the expectations for network capabilities and performance have significantly broadened \cite{giordani2020toward}. Beyond achieving higher data rates, \ac{6G} aims to support a wide range of advanced applications \cite{sahin2016flexible}, such as integrated artificial intelligence (AI) and communication, integrated sensing and communication (ISAC), and ubiquitous connectivity. Moreover, it seeks to enhance three existing scenarios: immersive communication, massive communication, and hyper-reliable low-latency communication. These scenarios are extensions of 5G’s established use cases, including \ac{eMBB}, \ac{mMTC}, and \ac{URLLC}, as defined by IMT-2030 \cite{IMT2030}.

Addressing these diverse requirements demands a highly adaptable \ac{PHY} capable of accommodating varying operational conditions, such as high-mobility environments, ultra-low latency, and the need for enhanced \ac{SE}. However, developing such a flexible \ac{PHY} introduces considerable challenges, particularly in waveform design.

\Ac{OFDM}, a foundational waveform for \ac{4G} and \ac{5G} systems, has been widely adopted due to its \ac{SE}, implementation simplicity, and compatibility with \ac{MIMO} systems \cite{4607239}. Despite these strengths, it has inherent drawbacks that become more pronounced as communication systems evolve. These include a high \ac{PAPR}, leading to power inefficiencies in uplink scenarios, particularly for energy-constrained devices, and poor performance in high-mobility environments, such as \ac{V2X} communication and high-speed railways \cite{noor20226g}, due to its sensitivity to Doppler shifts. Additionally, \ac{OFDM} requires strict synchronization and suffers from large sidelobes, limiting its effectiveness in applications requiring \ac{JSAC}.

To address these limitations, several waveform modifications were introduced in \ac{5G}. One prominent example is \ac{DFT-S-OFDM} \cite{sahin2016flexible}, which has been widely adopted for uplink transmission in \ac{5G} due to its ability to reduce \ac{PAPR}. \Ac{DFT-S-OFDM} retains the \ac{SE} of \ac{OFDM} while enhancing power efficiency, making it particularly well-suited for power-constrained devices. Additionally, variants such as \ac{ZT-DFT-S-OFDM} \cite{berardinelli2013zero} and \ac{UW-DFT-S-OFDM} \cite{sahin2015improved} were proposed to address out-of-band emissions and guard interval limitations, offering increased robustness against timing mismatches and multipath effects. While these advancements improved flexibility and performance, \ac{5G} waveforms still fail to meet the diverse requirements anticipated for \ac{6G} networks.

With the transition to \ac{6G}, the need for new waveforms capable of addressing high-mobility scenarios, doubly-dispersive channels, and ultra-low latency demands has become evident. Emerging candidates such as \ac{OTFS} \cite{hadani2017orthogonal}, \ac{AFDM} \cite{bemani2021afdm}, and \ac{OCDM} \cite{omar2021performance} have been designed to overcome the limitations of \ac{5G} waveforms in specific use cases. For instance, \ac{OTFS} operates in the \ac{DD} domain, offering exceptional resilience to Doppler effects, making it a strong candidate for high-mobility environments such as \ac{V2X} and low Earth orbit (LEO) satellite communications \cite{OTFS_high_mobility}. On the other hand, \ac{AFDM} demonstrates robustness in doubly-dispersive channels \cite{AFDM_doubly_dispersive} , where interference mitigation is critical. Similarly, \ac{OCDM}, leveraging chirp signals, excels in combating \ac{ISI}, making it highly suitable for scenarios demanding high \ac{SE} under challenging channel conditions.

While each of these waveforms offers unique benefits, their coexistence within a single network introduces new challenges. \Ac{6G} networks are expected to simultaneously support multiple waveforms, each tailored to specific applications with distinct performance requirements \cite{10382693}. For instance,  hyper reliable and low-latency communication demands ultra-low latency and high reliability, while massive  communication emphasizes massive connectivity and energy efficiency \cite{IMT2030}. This diversity necessitates a sophisticated framework capable of managing the orthogonal coexistence of these waveforms within shared time-frequency resources. Without effective coordination, waveform coexistence risks increased interference, suboptimal resource utilization, and degraded network performance.

The challenge extends beyond ensuring waveform compatibility to achieving high \ac{SE} and low implementation complexity. For \ac{6G} networks to meet their potential, a unified waveform framework is required—one capable of dynamically generating multiple waveforms and ensuring their orthogonal coexistence within the same spectral resources. This framework would enable the network to adapt to the diverse needs of \ac{6G} applications, allowing different waveforms to operate seamlessly without mutual interference. Moreover, such a unified approach is critical not only for supporting the expansive range of \ac{6G} use cases but also for maintaining backward compatibility with existing \ac{4G} and \ac{5G} systems.

This paper addresses a critical challenge in \ac{6G}—the need for a unified waveform architecture that enables the seamless and orthogonal coexistence of diverse waveforms, particularly those based on \ac{DFT} principles. We introduce the innovative concept of a "mother waveform," a foundational waveform designed to generate multiple specialized waveforms tailored for different applications, ensuring minimal interference and optimal resource utilization. This mother waveform framework not only accommodates \ac{6G} waveforms such as \ac{OTFS}, \ac{AFDM}, and \ac{OCDM}, but also maintains backward compatibility with legacy \ac{5G} waveforms like \ac{DFT-S-OFDM}, offering a cost-effective and efficient solution for the seamless transition to \ac{6G}. The key contributions of this paper are as follows:
\begin{itemize}
    \item This paper demonstrates that key waveforms used in 4G, 5G, and the emerging 6G networks, such as \ac{OFDM} and \ac{SC-IFDM}, are all based on the \ac{DFT}. We highlight how one of these \ac{DFT}-based waveforms can serve as a "mother waveform," capable of generating various other waveforms while maintaining efficiency and flexibility across generations.
    
    \item We propose an innovative framework that enables the derivation of multiple specialized "offspring" waveforms from the mother waveform. This framework offers enhanced flexibility in spreading techniques, resource allocation, and ensures backward compatibility, supporting seamless integration across different wireless standards.
    
    \item In addition, this paper introduces a novel flexible and sparse resource allocation scheme for \ac{OTFS} in the \ac{TF} domain, which allows a seamless coexistence scheme with other waveforms defined in the \ac{TF} domain.
    
    \item Furthermore, novel orthogonal coexistence schemes that enable sweeping-frequency waveforms (e.g., \ac{FMCW} \cite{4404963,8695699}) and constant-frequency waveforms (e.g., \ac{SC-IFDM} \cite{myung2006single,benvenuto2009single}, \ac{OTFS}) are proposed to operate simultaneously with minimal interference. This ensures efficient and interference-free communication within the same network, crucial for complex multi-waveform \ac{6G} environments.
    
    \item We conduct an in-depth, simulation-based analysis to assess the proposed coexistence schemes, focusing on key performance metrics such as complexity, \ac{BER}, system capacity, and sensing capabilities. These evaluations demonstrate the effectiveness of our approach across diverse network scenarios and applications.
\end{itemize}

The remainder of this article is organized as follows. Section \ref{sec:Pre} presents preliminaries about the existing waveforms in the literature. Section \ref{sec:DFTwaves} establishes the connection between SC-IFDM signal to different DFT waveforms. Section \ref{sec:proposedWaves} presents novel practical co-existence cases. In  Section \ref{sec:simulation}, the simulation results are provided and finally, Section \ref{sec:conclusion} concludes the paper.

\textit{Notation:} Bold uppercase $\mathbf{A}$, bold lowercase $\mathbf{a}$, and unbold letters $A,a$ denote matrices, column vectors, and scalar values, respectively. $(\cdot)^{-1}$ denote the inverse operators.  $\delta(\cdot)$ denotes the Dirac-delta function.  $\mathbb{C}^{{M\times N}}$ denotes the space of $M\times N$ complex-valued matrices.

\section{PRELIMINARIES} \label{sec:Pre}

\hspace{\parindent}In this section, we introduce the \ac{DZT}, followed by a discussion on key waveform models such as \ac{OFDM}, \ac{SC-IFDM}, \ac{OTFS}, \ac{FMCW}, \ac{OCDM}, and \ac{AFDM}.

\subsection{Discrete Zak Transform}
\hspace{\parindent}The \ac{DD} domain, also known as Zak space, is an intermediate space between the direct signal space "time" and the Fourier space "frequency". Assume the \ac{DD} grid is discretized into $M$ delays and $N$ Doppler bins, and consider an arbitrary discrete $MN$-point sequence $\boldsymbol{s}$. The \ac{DZT} of $\boldsymbol{s}$ is an isometry $\mathbb{C}^{MN} \rightarrow \mathbb{C}^{N \times M}$ expressed as \cite{DZT}
\begin{equation}
   \begin{aligned}
    X(k, l)= & \frac{1}{\sqrt{N}}\sum_{n=0}^{N-1} s(l+n M) e^{-j2 \pi  \frac{k}{N}n}, \\
    & \text{for}~ 0 \leq l<M, 0 \leq k<N .
\end{aligned} 
\label{equ:DZT}
\end{equation}

To compute $X(k,l)$, $M$ \ac{DFT} blocks of size $N$ are required for the data sets $s(l), s(l+M), \ldots, s(l+(N-1) M), ~ 0 \leq l<M$. The original sequence $\boldsymbol{s}$ can be recovered from its \ac{DZT} using the \ac{IDZT}, as follows
\begin{equation}
\begin{aligned}
         s(l+n M)&=\frac{1}{\sqrt{N}} \sum_{k=0}^{N-1} X(k,l) e^{j2 \pi \frac{k}{N}n} \\
        & \text{for}~ 0 \leq l<M, 0 \leq n<N .
\end{aligned}
\label{equ:IDZT}
\end{equation}

The vectorized output of the \ac{IDZT} for a matrix $\mathbf{X}$ can be written as 
\begin{equation}
    \mathbf{s}=\left(\mathbf{F}_{N}^\mathsf{H}\otimes\mathbf{I}_{M}\right)\text{vec}(\mathbf{X}),
\label{equ:final DZT}
\end{equation}
where $\mathbf{F}_N$ is $N \times N$ unitary \ac{DFT} matrix and $\mathbf{I}_{M}$ is the $M \times M$ identity matrix.


\subsection{OFDM Signal Model}

\hspace{\parindent}Consider a wireless communication system where $N$ data symbols are modulated using \ac{OFDM} over a total bandwidth $B$, operating at the carrier frequency $f_c$. The \ac{OFDM} modulator distributes the symbols $\{X(k),~k=0,\dots,N-1\}$ in the frequency domain. Denoting the bandwidth of each of the $N$ subcarriers as $\Delta f= B/N$, the time duration of one \ac{OFDM} symbol is $T = 1/\Delta f$. The frequency samples $X(k)$ are then converted to the time domain using an \ac{IDFT} operation. The discrete output of an \ac{OFDM} modulator is 
\begin{equation}
    s^{\text{OFDM}}(l)=\frac{1}{\sqrt{N}} \sum_{k=0}^{N-1} x^{\text{OFDM}}(k) e^{j 2 \pi \frac{k l}{N}},
\label{equ:ofdm}
\end{equation}
or equivalently in matrix form is given as
\begin{equation}
\mathbf{s}^{\text{OFDM}}= \mathbf{F}_{N}^\mathsf{H} \mathbf{x}^{\text{OFDM}}.
\end{equation}


\subsection{SC-IFDM Signal Model}
\hspace{\parindent}In \ac{SC-IFDM}, $M \times N$ data symbols are modulated by mapping them to $N$ \ac{DFT} blocks, each of size $M$. The outputs of these blocks are interleaved before applying a final $MN$-point \ac{IDFT}. The resulting time-domain signal $s^{\text{SC-IFDM}}(p)$, with the data symbols ${X^{\text{SC-IFDM}}(k,l)}$ for $k=0, \dots, N-1$ and $l = 0, \dots, M-1$, is given for $p=0,1,\dots,MN-1$ as \cite{chia2006distributed}
\begin{equation}
    s^{\text{SC-IFDM}}(p)=\frac{1}{\sqrt{N}} \sum_{k=0}^{N-1}  X^{\text{SC-IFDM}}(k,[p]_M)e^{j 2 \pi\frac{k}{MN}p}.
    \label{equ:simplify11}
\end{equation}

By setting $p=l+n M$ with $n=0,\dots,N-1$, we get
\begin{equation}
    s^{\text{SC-IFDM}}(l+n M)=\frac{1}{\sqrt{N}} \sum_{k=0}^{N-1} X^{\text{SC-IFDM}}(k,l) e^{j 2 \pi\frac{k(l+nM)}{MN}}.
    \label{equ:simplifypp}
\end{equation}

To express the discrete time-domain \ac{SC-IFDM} baseband signal in \eqref{equ:simplifypp} in a matrix-vector form, we write the output of the $M$-point \ac{DFT} block after interleaving as
\begin{equation}
    s(k+mN)=\frac{1}{\sqrt{M}} \sum_{l=0}^{M-1} X^{\text{SC-IFDM}}(k,l) e^{-j 2 \pi \frac{l}{M}m},
\label{equ:DZT_DFT}
\end{equation}
which is then written in matrix form as 
\begin{equation}
    \mathbf{s}=\mathbf{\Gamma} \left(\mathbf{I}_{N}\otimes\mathbf{F}_{M}\right)\mathbf{x}^{\text{SC-IFDM}},
\label{equ:DZT_DFT1}
\end{equation}
where $\mathbf{x}^{\text{SC-IFDM}}=\text{vec}(\mathbf{X}^{\text{SC-IFDM}})$, $\mathbf{\Gamma}=[\mathbf{\Gamma}_0, \dots, \mathbf{\Gamma}_{M-1}  ]^\text{T} $ is the interleaving matrix, where each $\mathbf{\Gamma}_m$ is defined as $\text{CircShift}\left((\mathbf{I}_N \otimes \mathbf{\Upsilon}),m\right)$ for $m=0, \dots, M-1$ and $\mathbf{\Upsilon}=[1,\mathbf{0}_{M-1}]$.
Thus, \eqref{equ:simplifypp} is reformulated as   
\begin{equation}
    \mathbf{s}^{\text{SC-IFDM}} =\mathbf{F}_{MN}^\mathsf{H}\mathbf{\Gamma}\left(\mathbf{I}_{N}\otimes\mathbf{F}_{M}\right)\mathbf{s}^{\text{SC-IFDM}}.
\label{equ:output}
\end{equation}


\subsection{OTFS Signal Model}

\hspace{\parindent}In \ac{OTFS}, $M \times N$ data symbols are modulated over a total bandwidth $B$ at the carrier frequency $f_c$. The \ac{OTFS} modulator maps the data symbols $\{X^{\text{OTFS}}(k,l),~k=0,\dots,N-1, ~ l = 0,\dots,M-1\}$ onto the \ac{DD} grid. Each subcarrier occupies a bandwidth of $\Delta f= B/M$, resulting in a time duration of $T = 1/\Delta f$. Consequently, an \ac{OTFS} frame spans a total duration of $NT$. The mapped symbols are transformed into the \ac{TF} domain using the \ac{ISFFT}, which is defined as
\begin{equation}
    X^{\mathrm{TF}}(n, m)=\frac{1}{\sqrt{M N}} \sum_{l=0}^{M-1} \sum_{k=0}^{N-1} X^{\text{OTFS}}(k,l) e^{j 2 \pi\left(\frac{n k}{N}-\frac{m l}{M}\right)}.
\label{equ:ISFFT}
\end{equation}

The time-domain \ac{OTFS} signal is then produced by converting the \ac{TF} samples to the time domain using the \ac{IDFT}, i.e., 
\begin{equation}
    s^{\text{OTFS}}(m+n M)=\frac{1}{\sqrt{M}} \sum_{m=0}^{M-1} X^{\mathrm{TF}}(n,m) e^{j 2 \pi \frac{m}{M} n}.
\label{equ:heisenberg}
\end{equation}

Note that the total length of the discrete sequences representing the frame is $M \times N$. Substituting \eqref{equ:ISFFT} in \eqref{equ:heisenberg} and using $(1 / M) \sum_{m=0}^{M-1} e^{j 2 \pi m \left(l-l^{\prime}\right)/ M}=\delta\left(l-l^{\prime}\right)$, we obtain
\begin{equation}
    s^{\text{OTFS}}(l+n M)=\frac{1}{\sqrt{N}} \sum_{k=0}^{N-1} X^{\text{OTFS}}(k,l) e^{j 2 \pi \frac{k}{N}n},
\label{equ:OTFS_IDZT}
\end{equation}
which is precisely the \ac{IDZT} described in \eqref{equ:IDZT}. 
As a result, the \ac{DFT}-based cascaded \ac{ISFFT} and \ac{OFDM} modulation have a similar structure to the \ac{IDZT}. In order to obtain the discrete sequence without employing the \ac{ISFFT}, one can use the \ac{IDZT}. Similar outcomes can be attained for the receiver when the \ac{DFT}-based \ac{OFDM} demodulation is coupled with the \ac{SFFT}, which yields the \ac{DZT}, i.e.,
\begin{equation}
    \mathbf{s}^{\text{OTFS}}=\left( \mathbf{F}_{N}^\mathsf{H}\otimes\mathbf{I}_{M}\right)\text{vec}(\mathbf{X}^{\text{OTFS}})=\left(\mathbf{F}_{N}^\mathsf{H}\otimes\mathbf{I}_{M} \right) \mathbf{x}^{\text{OTFS}}.
\label{equ:OTFS_IDZT_MTX}
\end{equation}


\subsection{FMCW Signal Model}

\hspace{\parindent}Consider a linear \ac{FMCW} signal, \( s^{\text{FMCW}}(t) \), sweeps linearly over a bandwidth \( B_c \) within a time duration \( T_c \). It is written in time-domain as \cite{csahin2020multi}
\begin{equation}
    s^{\text{FMCW}}(t) = e^{j\pi \eta t^2}, \quad 0 \leq t < T_c,
\label{equ:chirp}
\end{equation}
where $\eta = \frac{B_c}{T_c} $ is the chirp rate, and $\eta, T_c, B_c \in \mathbb{R}$. 
For a time-bandwidth product \( MN = \eta T_c^2 \) that is a positive integer, the FMCW signal can be critically sampled at intervals \( t = \frac{p}{MN} T_c \), leading to the discrete form given below
\begin{equation}
    s^{\text{FMCW}}(p) = e^{j\pi \frac{p^2}{MN}}, \quad 0 \leq p < MN.
\label{equ:discrete_chirp}
\end{equation}


\subsection{OCDM Signal Model}

\hspace{\parindent}In \ac{OCDM}, a set of $M \times N$ orthogonal chirps are used to modulate $M \times N$ data symbols $x$ using the \ac{IDFnT}. The discrete-time baseband \ac{OCDM} signal is expressed as \cite{ouyang2016orthogonal}

\begin{equation}
    s^{\text{OCDM}}(p)=\frac{e^{j \frac{\pi}{4}}}{\sqrt{MN}}\sum_{i=0}^{MN-1} x^{\text{OCDM}}(i)e^{-j \pi \frac{(p-i)^2}{MN}}
    \label{equ:OCDM},
\end{equation}
which can equivalently be expressed in matrix-vector form as
\begin{equation}
    \mathbf{s}^{\text{OCDM}}=\mathbf{\Phi}^{\mathsf{H}}  \mathbf{x}^{\text{OCDM}}, 
    \label{equ:OCDM_mtx}
\end{equation}
where $\mathbf{\Phi}$ is the unitary \ac{DFnT} matrix, defined with the $(p,i)^{th}$ elements as
\begin{equation}
    \Phi (p,i) = \frac{e^{j \frac{\pi}{4}}}{\sqrt{MN}} e^{-j \pi \frac{(p-i)^2}{MN}}, [MN]_2 =0.
    \label{equ:OCDM_elem}
\end{equation}

The matrix $\mathbf{\Phi}$ can be decomposed into a \ac{DFT} matrix with additional quadratic phase shifts
\begin{equation}
    \mathbf{\Phi} = \mathbf{\Theta}_1 \mathbf{F} \mathbf{\Theta}_2 ,
\end{equation}
where $\mathbf{\Theta}_1$ and $\mathbf{\Theta}_2$ are diagonal phase matrices given by $\mathbf{\Theta}_1 (p,p)=e^{-\frac{\pi}{4}}e^{j 2 \pi \frac{p^2}{MN}}$ and $\mathbf{\Theta}_1 (i,i)=e^{j 2 \pi \frac{p^2}{MN}}$.


\subsection{AFDM Signal Model}

\hspace{\parindent}In \ac{AFDM}, a vector of symbols $\mathbf{x}^{\text{AFDM}} \in \mathbb{C}^{MN \times 1}$ is mapped into the twisted \ac{TF} chirp domain using the \ac{IDAFT}, described by \cite{bemani2021afdm}
\begin{equation}
    s^{\text{AFDM}}(p)=\frac{1}{\sqrt{MN}}\sum_{i=0}^{MN-1}x^{\text{AFDM}}(i)e^{j2\pi \left(c_1 p^2 + c_2 i^2 + pi/MN\right)}, 
\label{equ:AFDM}
\end{equation}
where the $i$-th unmodulated element of the TF chirp domain is given by
\begin{equation}
    s^{\text{AFDM},i}(p)=e^{j2\pi \left(c_1 p^2 + c_2 i^2 + pi/MN\right)}.
\label{equ:one_AFDM}
\end{equation}

In a matrix form, \eqref{equ:AFDM} is represented as
\begin{equation}
\mathbf{s}^{\text{AFDM}}=\mathbf{\Lambda}_{c_1}^\mathsf{H} \mathbf{F}^\mathsf{H} \mathbf{\Lambda}_{c_2}^\mathsf{H}\mathbf{x}^{\text{AFDM}},
\label{equ:mtxAFDM}
\end{equation}
where $\mathbf{\Lambda}_{c_1}^\mathsf{H} \mathbf{F}^\mathsf{H} \mathbf{\Lambda}_{c_2}^\mathsf{H} \in \mathbb{C}^{MN \times MN}$ represents the forward $MN$-point \ac{DAFT} matrix, and the diagonal chirp matrix with a central digital frequency of $c_i$ represented by $\mathbf{\Lambda}_{c_i}  \triangleq \text{diag}[e^{-j2\pi c_i (0)^2}, \dots,e^{-j2\pi c_i (MN-1)^2}$. 
This structure shows that \ac{AFDM} can also be classified as a \ac{DFT}-based waveform due to its use of chirp modulation.
\section{SC-IFDM To different DFT waveforms} \label{sec:DFTwaves}

\hspace{\parindent}In this section, we demonstrate that all discussed waveforms (OFDM, OTFS, FMCW, OCDM, AFDM) can be efficiently generated using the general structure of \ac{SC-IFDM}.


\subsection{SC-IFDM to OFDM}

\hspace{\parindent}When \ac{SC-IFDM} is reduced to a single-dimensional waveform by setting $M=1$ in \eqref{equ:simplifypp}, the result is the following expression
\begin{equation}
    s_{\text{SC-IFDM}}^{\text{OFDM}}(l)=\frac{1}{\sqrt{N}} \sum_{k=0}^{N-1} x^{\text{OFDM}}(k) e^{j 2 \pi\frac{k}{N}n } ,
    \label{equ:DFT_OFDM}
\end{equation}
which exactly matches the \ac{OFDM} signal, as described in \eqref{equ:ofdm}.


\subsection{SC-IFDM to OTFS}
\begin{figure}[t]
   \centering
\includegraphics[width=0.47\textwidth]{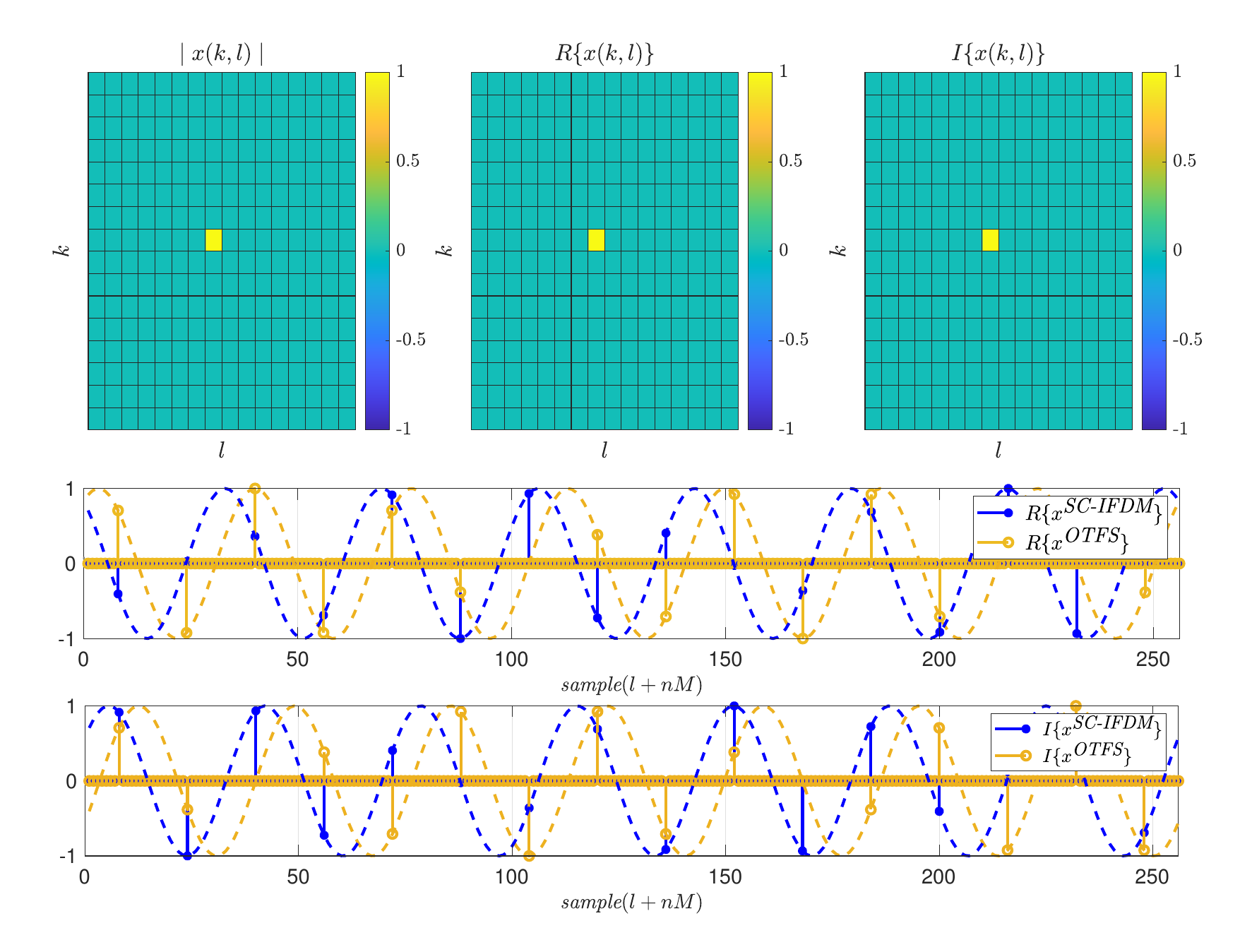}
    \caption{SC-IFDM to OTFS, $M=16$, $N=16$.}
    \label{fig:OTFSvsDFT}
\end{figure}

\hspace{\parindent}By analyzing \eqref{equ:simplifypp}, it becomes clear that \ac{SC-IFDM} and \ac{OTFS} exhibit similar data-spreading behavior, with the primary difference being the phase shifts applied in the time and the \ac{DD} domains. In \ac{OTFS} modulation, each delay bin contains $M$ repetitions of the Doppler domain data. This repetition can be achieved by performing an $N$-point \ac{IDFT} followed by upsampling the output by $M$ samples, which is equivalent to taking an $MN$-point \ac{IDFT}. The \ac{OTFS} time domain signal in \eqref{equ:OTFS_IDZT} can be represented in terms of \ac{SC-IFDM} as 
\begin{equation}
\begin{aligned}
    X_{\text{SC-IFDM}}^{\text{OTFS}}(k,l)&= \frac{1}{\sqrt{N}} \sum_{n=0}^{N-1}s^{\text{OTFS}}(l+n M)e^{-j2\pi\frac{k(l+nM)}{MN} }\\
    & =X^{\text{OTFS}}(k,l)\omega_{k}^{l},
\label{equ:otfs_dft}
\end{aligned}
\end{equation}
where $\omega_{k}^{l} = e^{j 2 \pi (\frac{-kl}{MN})}$. Hence, the difference between \ac{OTFS} and \ac{SC-IFDM} is merely a phase shift, as illustrated in Fig. \ref{fig:OTFSvsDFT}. 
Both waveforms share similar structures, and \ac{OTFS} can be generated from \ac{SC-IFDM} by applying phase-adjustments to the same data. The matrix form of \eqref{equ:otfs_dft} is written as
\begin{equation}
    \mathbf{s}_{\text{SC-IFDM}}^{\text{OTFS}}=\mathbf{F}_{MN}^\mathsf{H}\mathbf{\Gamma} \left(\mathbf{I}_{N}\otimes\mathbf{F}_{M}\right)\mathbf{\Psi} \mathbf{x^{\text{OTFS}}},
\end{equation}
where $\mathbf{\Psi}=\text{diag}\{\boldsymbol{\omega}_0, \dots, \boldsymbol{\omega}_{N-1}\}$ and $\boldsymbol{\omega}_k=[\omega_k^0, \dots, \omega_{k}^{M-1}]^\text{T}$, with $\omega_k^l=e^{-j2\pi\frac{kl}{MN}}$. This allows \ac{SC-IFDM} to experience the same delay and Doppler shifts as \ac{OTFS} under doubly dispersive channels, and their behavior in the effective channel is equivalent. The received \ac{SC-IFDM} signal, $\mathbf{y^{\text{SC-IFDM}}}$, can be written as 
\begin{equation}
\begin{aligned}
    \mathbf{y^{\text{SC-IFDM}}}=\mathbf{H} \mathbf{F}_{MN}^\mathsf{H}\mathbf{\Gamma} \left(\mathbf{I}_{N}\otimes\mathbf{F}_{M}\right) \mathbf{x}^{\text{SC-IFDM}}+ \mathbf{w},
\end{aligned}
\end{equation}

where $\mathbf{H}$ represents the $R$-tap time-domain channel matrix\footnote{The delay and Doppler shifts are assumed to be integers.}, and $\mathbf{w}$ denotes the \ac{AWGN} vector with variance $\sigma^2$. Specifically, $\mathbf{H}$ is defined as
\begin{equation}
    \mathbf{H} = \sum_{r=0}^{R-1} h_{r} \boldsymbol{\Pi}^{l_{r}}_{MN} \boldsymbol{\Delta}^{k_{r}}_{MN},
    \label{equ:Channel_gen}
\end{equation}
where $l_r$, $k_r$, and $h_r$ denote the delay, Doppler shifts, and channel coefficients of the $r$-th path, respectively. The matrix $\boldsymbol{\Pi}_{MN}^{l_r}$ is a permutation matrix with elements $\Pi_{MN}^{l_r}(k, l) = \delta([l-k]_{MN} - l_r)$, and $\boldsymbol{\Delta}_{MN}^{k_{r}}$ is an $M N \times M N$ diagonal matrix defined as
\begin{equation}
    \boldsymbol{\Delta}_{MN}^{k_{r}} = \operatorname{diag}\left[z_r^{0}, z_r^{1}, \ldots, z_r^{MN-1}\right],
\end{equation}
where $z_r = e^{\frac{j 2 \pi k_{r}}{M N}}$.
Thus, the transformed received \ac{SC-IFDM} signal, $\mathbf{\hat{Y}}^{\text{SC-IFDM}}$, can be expressed as
\begin{equation}
\begin{aligned}       
    \mathbf{\hat{Y}}^{\text{SC-IFDM}}=&\underbrace{\left(\mathbf{I}_{N}\otimes\mathbf{F}_{M}^\mathsf{H}\right)\mathbf{\Gamma}^\mathsf{H}\mathbf{F}_{MN}
    \mathbf{H}\mathbf{F}_{MN}^\mathsf{H}\mathbf{\Gamma} \left(\mathbf{I}_{N}\otimes\mathbf{F}_{M}\right)}_{\mathbf{H_{\text{eff}}^{\text{SC-IFDM}}}}\\& \mathbf{X}^{\text{SC-IFDM}}+ \left(\mathbf{I}_{N}\otimes\mathbf{F}_{M}^\mathsf{H}\right)\mathbf{\Gamma}^\mathsf{H} \mathbf{F}_{MN}\mathbf{w} ,
\end{aligned}
\end{equation}
while the received and transformed \ac{OTFS} signal can be written as 
\begin{equation}
\begin{aligned}       
    \mathbf{\hat{Y}_{\text{SC-IFDM}}^{\text{OTFS}}}&=\underbrace{\mathbf{\Psi}^{\mathsf{H}}\mathbf{H_{\text{eff}}^{\text{SC-IFDM}}}\mathbf{\Psi} }_{\mathbf{H_{\text{eff}}^{\text{OTFS}}}}  \mathbf{x^{\text{OTFS}}}\\&+ \mathbf{\Psi}^{\mathsf{H}}\left(\mathbf{I}_{N}\otimes\mathbf{F}_{M}^\mathsf{H}\right)\mathbf{\Gamma}^\mathsf{H} \mathbf{F}_{MN}\mathbf{w}.
\end{aligned}
\end{equation}

Therefore, \ac{OTFS} and \ac{SC-IFDM} can achieve similar performance under doubly dispersive channels, as one can be derived from the other through appropriate phase adjustments. Additionally, an \ac{SC-IFDM} receiver can extract \ac{OTFS} data, and vice versa.


\subsection{SC-IFDM to FMCW}

\begin{figure}[t]
   \centering
\includegraphics[width=0.47\textwidth]{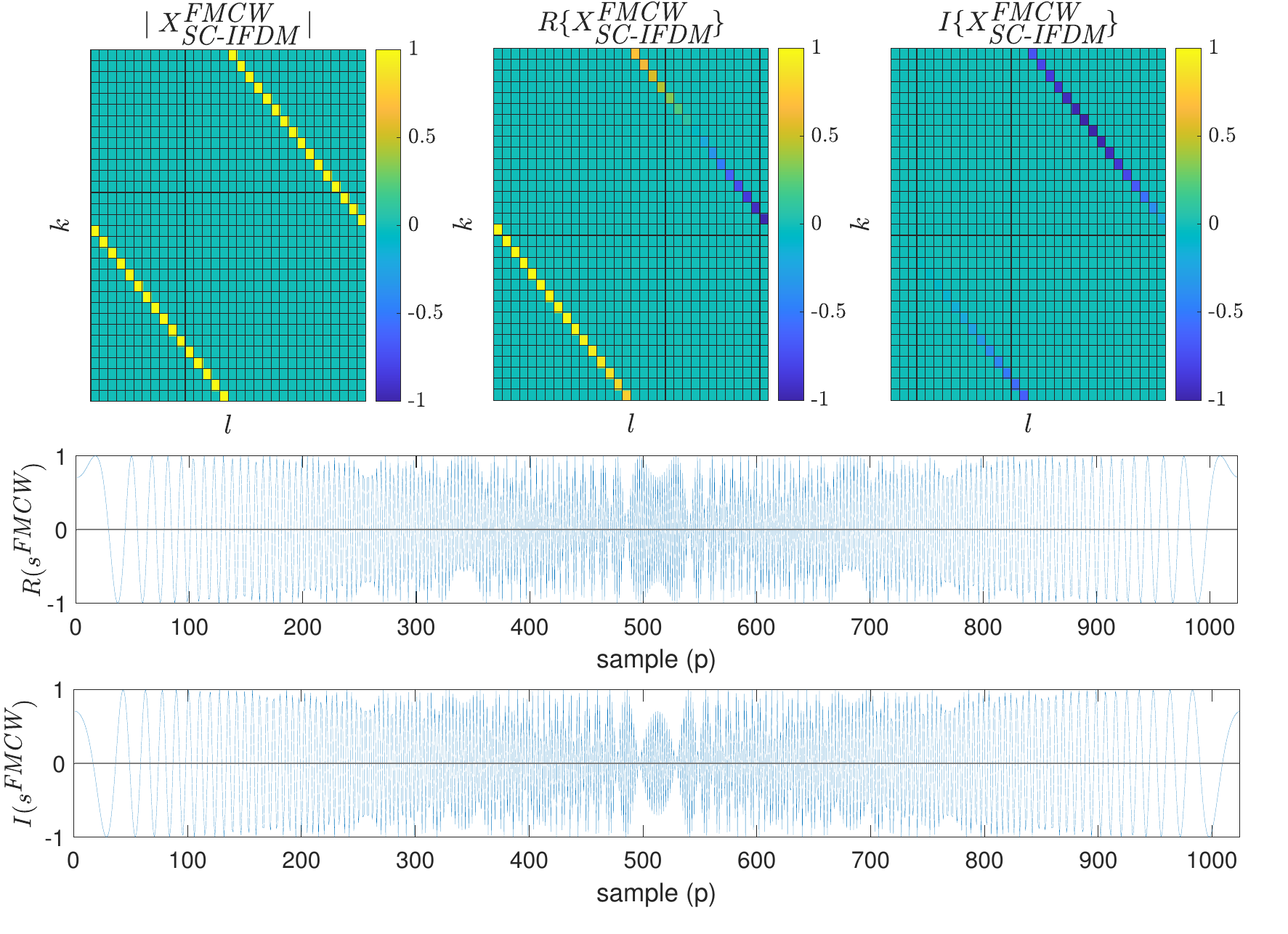}
    \caption{ SC-IFDM to FMCW,  $M=32,N=32.$ }
    \label{fig:ChirpvsDFT}
\end{figure}

\hspace{\parindent}Taking $p=l+n M$, the discrete \ac{FMCW} signal in \eqref{equ:discrete_chirp} can be expressed in the \ac{SC-IFDM} domain using the inverse transform of \eqref{equ:simplifypp}, i.e.,
\begin{equation}
\begin{aligned}
    X_{\text{SC-IFDM}}^{\text{FMCW}}(k,l)&= \frac{1}{\sqrt{N}} \sum_{n=0}^{N-1}s^{\text{FMCW}}(l+n M)e^{-j2\pi\frac{k(l+nM)}{MN} }\\
    &=\frac{e^{j\pi (\frac{l^2}{MN}  ) }\omega_{k}^{l}}{\sqrt{N}} \sum_{n=0}^{N-1}e^{j2\pi\frac{\left(\frac{M}{2} n^2+(l-k) n\right)}{N} }.
\label{equ:chirp_dft_1}
\end{aligned}
\end{equation} 

Note that \( e^{j\pi n} = 1 \) for even \( n \) and \( -1 \) for odd \( n \). Similarly, \( n^2 \) is even when \( n \) is even and odd when \( n \) is odd. Given that \( M \in \mathbb{Z} \), it follows that $e^{j\pi (M n^2)} = (-1)^{M n^2} = (-1)^{M n}$. Incorporating this result into \eqref{equ:chirp_dft_1}, we obtain
\begin{equation}
    X_{\text{SC-IFDM}}^{\text{FMCW}}(k,l)=\frac{e^{j\pi (\frac{l^2}{MN}  ) }\omega_{k}^{l}}{\sqrt{N}} \sum_{n=0}^{N-1}e^{j2\pi\frac{\left(\frac{M}{2} +l-k\right) n}{N} }.
\end{equation}

The elements of the matrix  $X_{\text{SC-IFDM}}^{\text{FMCW}}(k,l)$ are nonzero only when $\left[\frac{M}{2} +l-k\right]_N=0$, leading to a sparsity condition where the matrix is nonzero for certain values of $l$, as depicted in Fig. \ref{fig:ChirpvsDFT}.  Specifically, \( X_{\text{SC-IFDM}}^{\text{FMCW}}(k,l) \) simplifies to \( s^{\text{FMCW}}(l) \omega_{k}^{l} \) for the valid solutions corresponding to each \( l = 0, \dots, M-1 \). This results in a sparse matrix representation, where only $M$ nonzero elements exist in the \ac{DFT}-based space of the \ac{SC-IFDM}.
Thus, \ac{FMCW} signaling can be efficiently generated using \ac{SC-IFDM} by selecting the first $M$ samples from $\mathbf{s}^{\text{FMCW}}$ based on this sparsity condition. This approach offers a computationally efficient method to integrate the advantages of \ac{FMCW} and \ac{SC-IFDM} for \ac{JSAC} systems.
\subsection{SC-IFDM to OCDM}

\hspace{\parindent}The $i$-th \ac{OC} of \ac{OCDM} in \eqref{equ:OCDM} can be expressed as
\begin{equation}
    s^{\text{OCDM},i}(l+n M)=e^{j \frac{\pi}{4}} e^{-j \pi \frac{\left((l+nM)-i \right)^2}{MN}}.
\label{equ:kth_Disc_chirp2}
\end{equation}

Mapping $x^{\text{OCDM},i}$ into the \ac{SC-IFDM} domain gives
\begin{equation}
\begin{aligned}
    &X_{\text{SC-IFDM}}^{\text{OCDM},i}(k,l)= \frac{1}{\sqrt{N}} \sum_{n=0}^{N-1}s^{\text{OCDM},i}(l+n M)e^{-j 2 \pi\left( \frac{k(l+nM)}{MN}\right)}\\
    & = \frac{e^{j\frac{\pi}{4}}e^{j\pi (\frac{(l-i)^2}{MN}  ) }\omega_{k}^{l}}{\sqrt{N}} \sum_{n=0}^{N-1}e^{j2\pi\frac{\left(\frac{-M}{2} n^2+(i-l-k) n\right)}{N} }.
\label{equ:ocdm_dft}
\end{aligned}
\end{equation}

\begin{figure}[t]
   \centering
\includegraphics[width=0.48\textwidth]{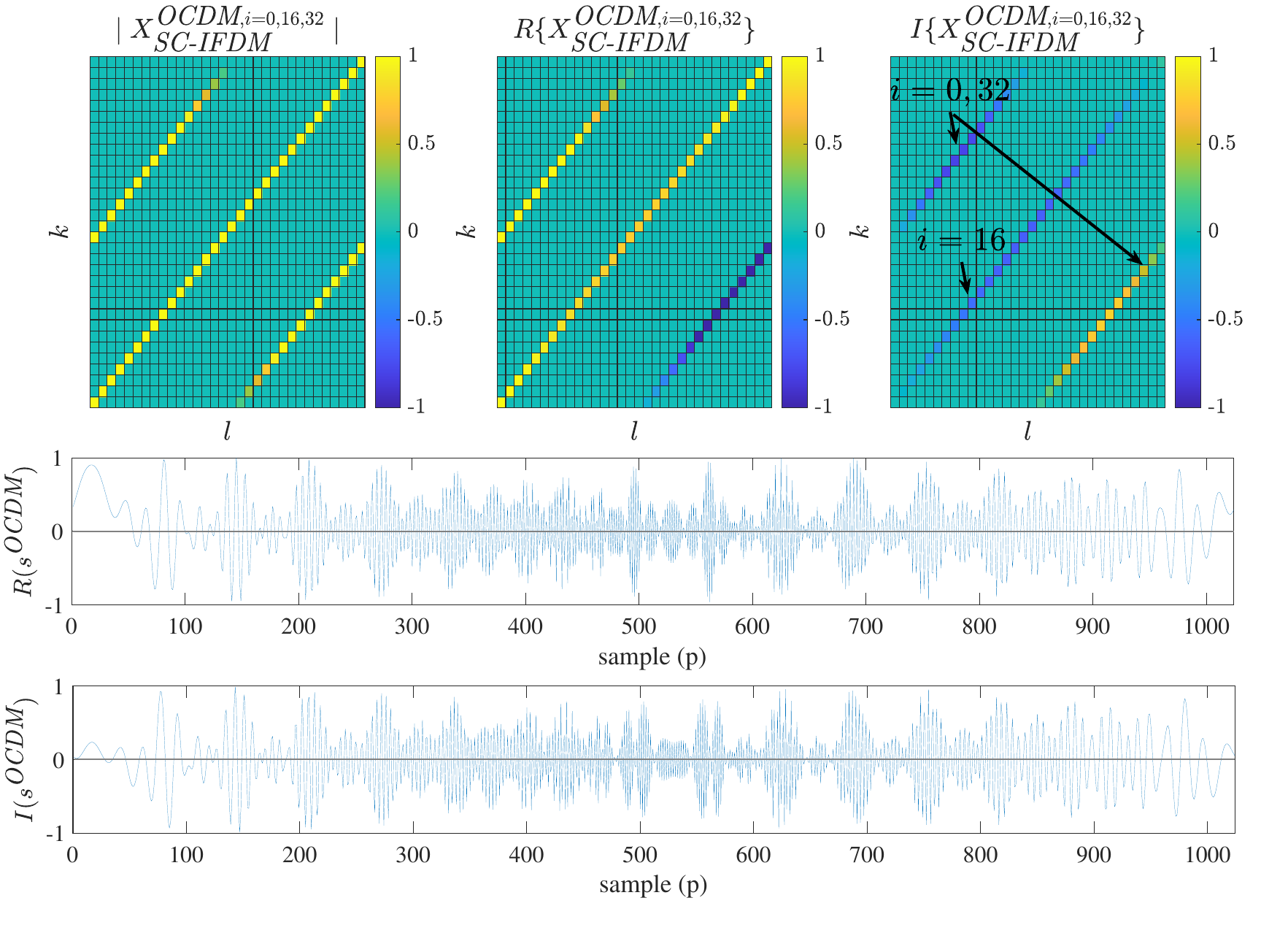}
    \caption{SC-IFDM to OCDM, $M=32,N=32$. }
    \label{fig:OCDMvsDFT}
\end{figure}

The sparsity of \ac{OCDM} in \ac{SC-IFDM} is apparent, as nonzero elements occur only at \( M \) specific points, as illustrated in Fig.~\ref{fig:OCDMvsDFT}. Consequently, \ac{OCDM} signals can be efficiently generated within the \ac{SC-IFDM} framework by leveraging this sparse representation. To demonstrate this, \eqref{equ:ocdm_dft} can be rewritten as

\begin{equation}
    X_{\text{SC-IFDM}}^{\text{OCDM},i}(k,l)=\frac{e^{-j\frac{\pi}{4}}e^{j\pi (\frac{(l-i)^2}{MN}  ) }\omega_{k}^{l}}{\sqrt{N}}
    \sum_{n=0}^{N-1}e^{j2\pi\frac{\left(\frac{-M}{2}+i-l-k\right)n}{N} }.
\end{equation}

For each $i$-th element, the condition $\left[\frac{-M}{2}+i-l-k\right]_N=0$ has a single solution for each $l=0,\dots,M-1$. 
For each $[i]_N$, the \acp{OC} overlap at the same $l,k$ indices, which demonstrates that \ac{OCDM} is sparse and represented by only $M$ nonzero points in \ac{SC-IFDM} space, as shown in Fig. \ref{fig:OCDMvsDFT}.  
An \( M \times N \) \ac{OCDM} signal consists of \( M \) non-overlapping orthogonal signals, such as for \( i=16 \) (occupying distinct indices), each containing \( N \) overlapping orthogonal signals in \ac{SC-IFDM} on the same \( k,l \) indices, such as for \( i=0 \) and \( i=32 \) (occupying the same indices while maintaining orthogonality). 
\subsection{SC-IFDM to AFDM}
\hspace{\parindent}Taking $p=l+nM$ in \eqref{equ:one_AFDM}, $i$-th \ac{AFDM} chirp element can be mapped to the \ac{SC-IFDM} as 
\begin{equation}
\begin{aligned}
    &X_{\text{SC-IFDM}}^{\text{AFDM},i}(k,l)= \frac{1}{\sqrt{N}} \sum_{n=0}^{N-1}s^{\text{AFDM},i}(l+n M)e^{-j 2 \pi\left( \frac{k(l+nM)}{MN}\right)}\\
    & = \frac{s^{\text{AFDM},i}(l) \omega_{k}^{l}}{\sqrt{N}}\sum_{n=0}^{N-1}e^{j2\pi\frac{\left(c_1 NM^2 n^2+(i+c_1 2MN l-k) n\right)}{N} }.
  \label{equ:afdm_dft}
  \end{aligned}
\end{equation}

By setting $c_1=\frac{c_1^{'}}{2MN}$, \eqref{equ:afdm_dft} can be given as  
\begin{equation}
    X_{\text{SC-IFDM}}^{\text{AFDM},i}(k,l)=\frac{s^{\text{AFDM},i}(l) \omega_{k}^{l}}{\sqrt{N}}
    \sum_{n=0}^{N-1}e^{j2\pi\frac{\left(c_1^{'}\frac{M}{2}+c_1^{'}l+ i-k \right)n}{N} }.
  \label{equ:afdm_dft_1}
\end{equation}

\begin{figure}[t]
   \centering
\includegraphics[width=0.479\textwidth]{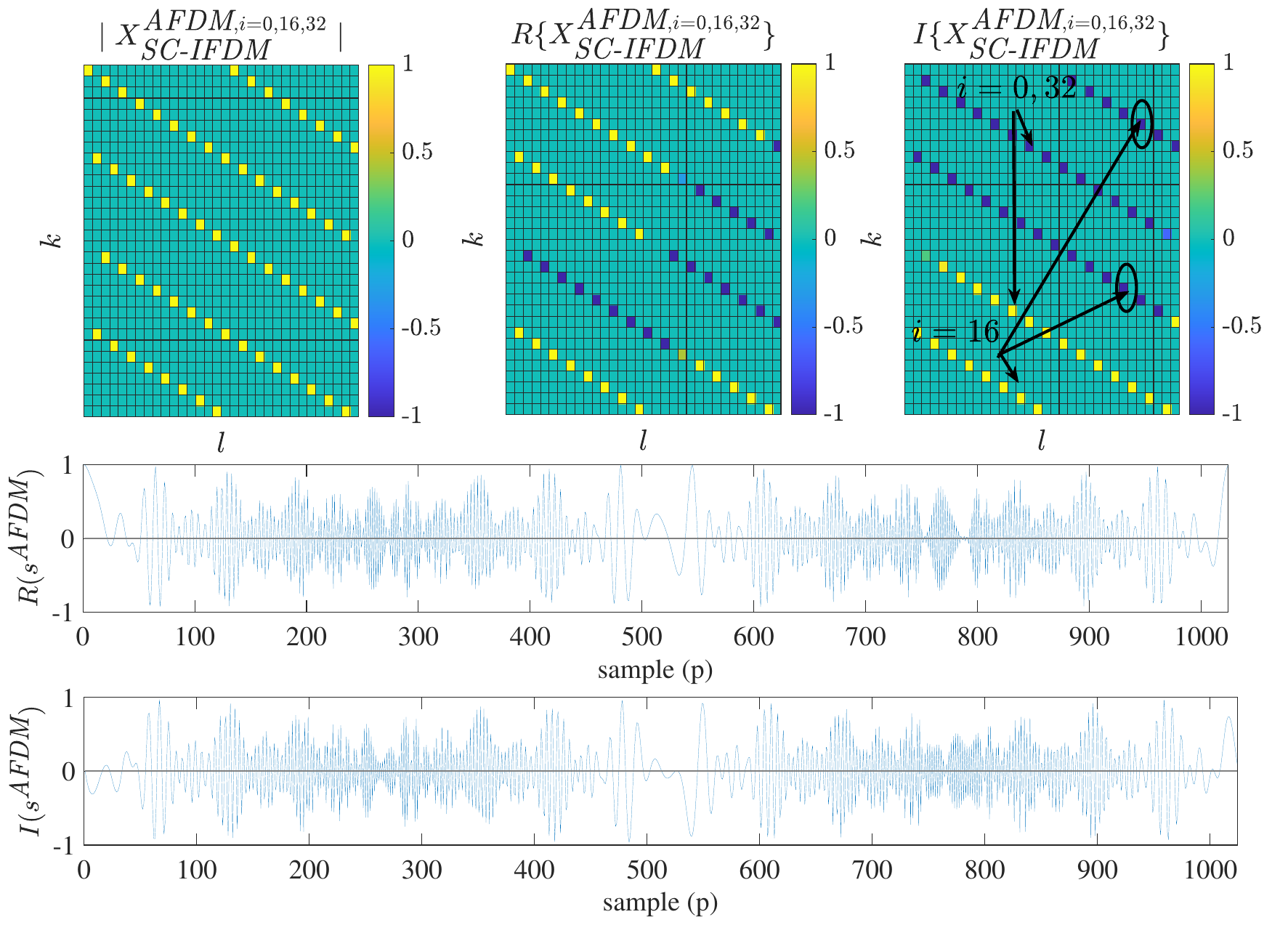}
    \caption{ SC-IFDM to AFDM $M=32,N=32, c_1^{'} = 2, c_2=1.$}
    \label{fig:AFDMvsDFT}
\end{figure}

Therefore, \ac{AFDM} is sparse, and \( X_{\text{SC-IFDM}}^{\text{AFDM},i}(k,l) \) is nonzero only if $\left[c_1^{'}\frac{M}{2}+c_1^{'}l+ i-k\right]_N=0$, i.e., for each  $i$-th chirp, there are only \( M \) solutions with a more sparse representation where, for each \( l \), there are \( c_1^{'} \) solutions. Thus, \ac{AFDM} can be represented by only \( M \) nonzero points in the \ac{SC-IFDM} space, as illustrated in Fig.~\ref{fig:AFDMvsDFT}. For each $[i]_N$, the \ac{AFDM} twisted chirps overlap at the same $l$ and $k$ indices.
Hence, the \ac{AFDM} signal can be generated using \ac{SC-IFDM}. 
Notably, the element \( X^{\text{SC-IFDM}}(k,l) \) exhibits a spreading pattern over the $[i]_N$indices in the twisted \ac{TF} chirp domain of \ac{AFDM}.  As a result, \ac{SC-IFDM} is also sparse in the twisted chirp domain, allowing \ac{AFDM} to be efficiently generated within the \ac{SC-IFDM} structure.

\begin{figure*}[t]
   \centering
\includegraphics[width=0.97\textwidth]{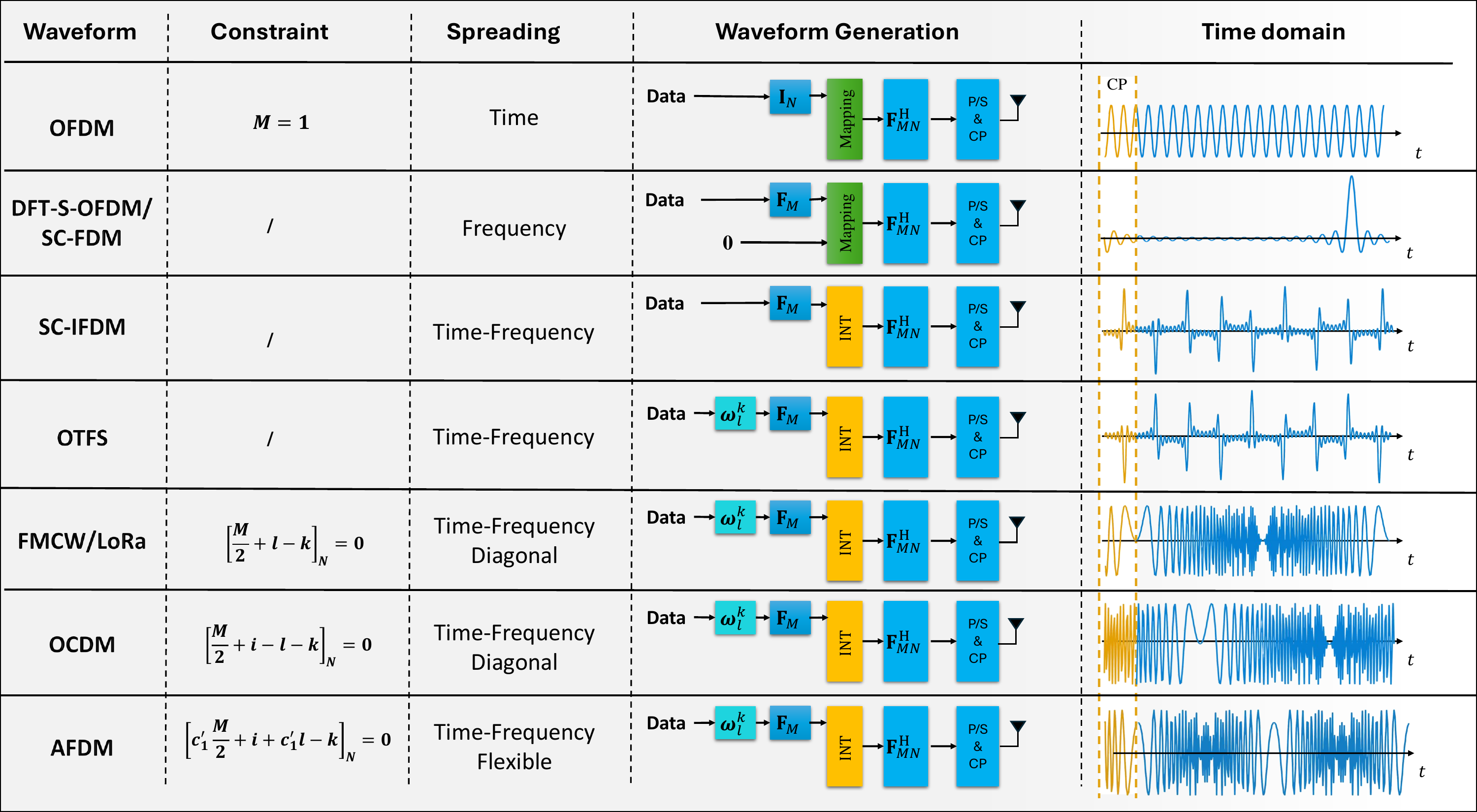}
    \caption{SC-IFDM as a mother waveform to all DFT waveforms and illustration of the resulting signals in 
time domain.}
    \label{fig:Mother_waveform}
\end{figure*}
\subsection{Proposed Mother Waveform}
\hspace{\parindent}We have shown that \ac{OFDM}, \ac{OTFS}, \ac{FMCW}, \ac{OCDM}, and \ac{AFDM} can be efficiently generated using the unified \ac{SC-IFDM} framework. This insight extends to other advanced waveforms, such as \ac{GFDM}, \ac{ZT-OFDM}, \ac{UW-OFDM}, Enhanced \ac{UW-OFDM}, and UW DFT-S-W-OFDM, which share similar structural foundations, as noted in \cite{sahin2016flexible}. These structural commonalities inspire the development of a generalized "mother waveform" based on \ac{SC-IFDM}, from which all other waveforms can be derived through simple phase adjustments and customized data allocation techniques. The primary advantage of this approach lies in its backward compatibility with existing systems, requiring no hardware modifications. This framework capitalizes on current infrastructure, offering a cost-effective pathway to introducing new waveforms without the need for substantial upgrades. By generating all \ac{DFT}-based waveforms from a single versatile structure, the proposed framework provides a dynamic \ac{PHY} that can adapt to a wide range of applications, offering flexibility in both waveform design and resource allocation. 

As illustrated in Fig. \ref{fig:Mother_waveform}, 
this mother waveform framework systematically generates various waveforms by adjusting how data blocks are allocated, either through interleaving or direct mapping. It allows targeted waveform creation through specific mappings in the grid $k$ and $l$ indices. This flexibility enables seamless orthogonal coexistence of multiple waveforms, allowing \ac{4G}, \ac{5G}, and \ac{6G} systems to support diverse applications with minimal interference.
Moreover, this approach enhances the scalability and adaptability of communication systems, enabling the coexistence of multiple waveforms within the same \ac{TF} resources. It offers a robust solution that meets the evolving demands of future wireless networks, balancing high performance with the seamless integration of heterogeneous services. By facilitating smooth transitions between generations and efficiently managing spectrum resources, the proposed framework ensures that both current and future applications can coexist without requiring changes to the existing hardware infrastructure, making it a highly scalable and future-proof solution for next-generation networks.

\section{Practical Co-existence Strategies of The Proposed Approach for 6G} \label{sec:proposedWaves}
\hspace{\parindent}To emphasize the effective integration across the \ac{TF} domain discussed in the previous section, this work explores orthogonal co-existence capabilities between \ac{OTFS} or \ac{SC-IFDM} and \ac{OFDM}, as well as between \ac{SC-IFDM} and \ac{AFDM}.
First, we introduce a novel resource allocation strategy for \ac{OTFS}, designed to facilitate its orthogonal co-existence with other waveforms. Second, we discuss a specific scenario demonstrating orthogonal co-existence between \ac{OTFS} and \ac{OFDM}. Lastly, we propose a method for achieving orthogonal co-existence between \ac{SC-IFDM} and \ac{AFDM}.
These discussions highlight the effectiveness of the mother waveform concept in supporting orthogonal co-existence while demonstrating its versatility to extend beyond the presented cases for diverse applications.

\subsection{OTFS and SC-IFDM Resource Allocation Scheme}

\hspace{\parindent}Both \ac{OTFS} and \ac{SC-IFDM} distribute data across the entire \ac{TF} domain to capture delay and Doppler information. However, this inherently prevents orthogonal coexistence with other waveforms within the same resources. Additionally, when an \ac{OTFS} or \ac{SC-IFDM} user requires less data than the total grid, the data must still occupy the entire \ac{TF} resources to maintain delay and Doppler resolution, leading to reduced \ac{SE}.
To address this limitation, we propose a method to optimize resource allocation in the \ac{TF} domain without compromising the delay and Doppler representation or resolution. This approach leverages the inherent capability of the \ac{OTFS} system to map \ac{TF} resources using the \ac{ISFFT} transform. The delay and Doppler resolution remain dependent solely on the system's bandwidth and symbol duration, ensuring efficient and flexible allocation.

Consider an \ac{OTFS} system with $\frac{MN}{\alpha \beta}$ data bins, represented as $x[k',l']$, where $k' = 0, \dots, \frac{N}{\alpha}-1$ and $l'=0, \dots, \frac{M}{\beta}-1$. Our approach employs a Kronecker product with a precoding matrix to precisely allocate resources in the \ac{TF} domain, mathematically expressed as
\begin{equation}
    \mathbf{X}_{pr}= \mathbf{P}\otimes \mathbf{X},
    \label{equ:kron}
\end{equation}
where the precoding matrix is defined as
\begin{equation}
\begin{aligned}
    \mathbf{P}_{q_1,q_2} = \mathbf{f}_{q_1}^\mathsf{H} \mathbf{f}_{q_2}.
    \label{equ:precoding_matrix}
\end{aligned}
\end{equation}

Here, $\mathbf{f}_{q_1}^\mathsf{H}$ denotes the conjugate of the $q_1$-th column of the $\alpha \times \alpha$ \ac{DFT} matrix, and $\mathbf{f}_{q_2}$ represents the $q_2$-th row of the $\beta \times \beta$ \ac{DFT} matrix. The matrix in \eqref{equ:precoding_matrix} can be written explicitly as
\begin{equation}
\begin{aligned}
    P_{q_1,q_2}[a,b] = \frac{1}{\sqrt{\alpha \beta}}e\left\{\frac{-a q_1}{\alpha}+\frac{b q_2}{\beta}\right\},
\end{aligned}
\end{equation}
where $e\{\cdot\}=e^{j 2 \pi (\cdot)}$, $a=\lfloor{\frac{k\alpha}{N}}\rfloor$, and $b=\lfloor{\frac{l\beta}{M}}\rfloor$. Therefore, the precoded signal, $\mathbf{X}_{pr}$, is expressed as
\begin{equation}
    {X}_{\text{pr}}[k,l] = \frac{1}{\sqrt{\alpha \beta}}X_s([k]_{\frac{N}{\alpha}},[l]_{\frac{M}{\beta}})e\left\{\frac{-a q_1}{\alpha}+\frac{b q_2}{\beta}\right\}.
    \label{equ:precoded_OTFS}
\end{equation}

By substituting $k=k'+a\frac{N}{\alpha}$ and $l=l'+b\frac{M}{\beta}$ into \eqref{equ:ISFFT}, where $k'=0, \dots, \frac{N}{\alpha}-1$ and $l'=0, \dots, \frac{M}{\beta}-1$, the \ac{TF} domain signal is given by
\begin{equation}
\begin{aligned}
    &X_{\text{TF}}[n,m] = \frac{1}{\sqrt{MN\alpha \beta}} \sum_{a=0}^{\alpha-1} \sum_{b=0}^{\beta-1} \sum_{k'=0}^{\frac{N}{\alpha}-1} \sum_{l'=0}^{\frac{M}{\beta}-1} {x}_{s}[k', l'] \\
    &\times e\left\{\frac{n(k'+\frac{aN}{\alpha})}{N}-\frac{m(l'+\frac{bM}{\beta})}{M}\right\} e\left\{\frac{b q_2}{\beta}-\frac{a q_1}{\alpha}\right\}.
\end{aligned}
    \label{equ:TF1}
\end{equation}

After simplification, $X_{\text{TF}}[n,m]$ can be written as
\begin{equation}
\begin{aligned}
    &X_{\text{TF}}[n,m] = \frac{\psi}{\sqrt{\alpha \beta}} \sum_{a=0}^{\alpha-1} \sum_{b=0}^{\beta-1} e\left\{\frac{a(n-q_1)}{\alpha}-\frac{b(m-q_2)}{\beta}\right\},
\end{aligned}
    \label{equ:TF2}
\end{equation}
where $\psi = \frac{1}{\sqrt{MN}} \sum_{k'=0}^{\frac{N}{\alpha}-1} \sum_{l'=0}^{\frac{M}{\beta}-1} x_s[k',l']e\left\{\frac{nk'}{N}-\frac{ml'}{M}\right\}$. To further simplify $X_{\text{TF}}[n,m]$, \eqref{equ:TF2} can be expressed as a product of two Kronecker delta functions as follows
\begin{equation}
\begin{aligned}
    &X_{\text{TF}}[n,m] = \psi \delta_{nq_1} \delta_{mq_2} = \psi \sum_{a=0}^{\alpha-1} e^{\left\{ \frac{a(n-q_1)}{\alpha}\right\}} \sum_{b=0}^{\beta-1} e^{\left\{ \frac{b(q_2-m)}{\beta}\right\}}.
\end{aligned}
\label{equ:kron1}
\end{equation}

For $X_{\text{TF}}[n,m]$ to be nonzero, both $\delta_{nq_1}$ and $\delta_{mq_2}$ must be nonzero, which occurs only when $[n-q_1]_{\alpha}=0$ and $[m-q_2]_{\beta}=0$. Thus, the \ac{TF} grid for $X_{\text{TF}}[n,m]$ can be controlled as
\begin{equation}
\begin{aligned}
    X_{\text{TF}}[n,m] =
    \begin{cases}
    \psi, & [n-q_1]_{\alpha}=0\, \text{and} \, [m-q_2]_{\beta}=0, \\
    0, & \text{Otherwise}.
    \end{cases}.
\end{aligned}
\end{equation}

\begin{figure}[t] 
    \includegraphics[width=0.47\textwidth]{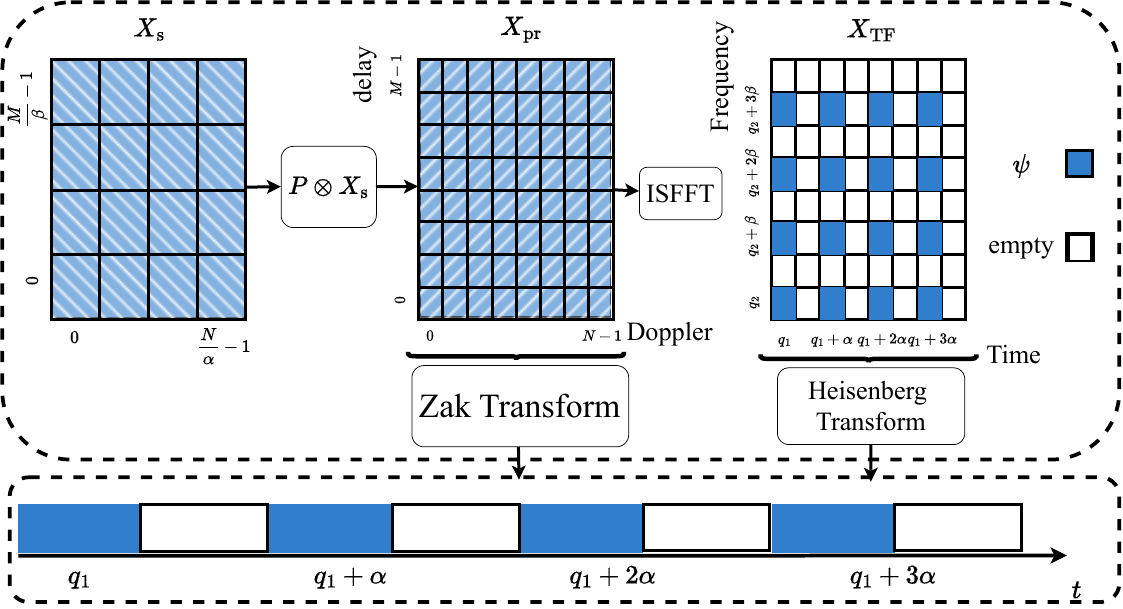}
    \caption{Proposed OTFS resource allocation ($\alpha=2, \beta=2$). }
    \label{fig:Resource allocation}
\end{figure}

This demonstrates that the \ac{OTFS} grid in \ac{TF} domain can be controlled by adjusting the precoding matrix $P_{q_1,q_2}[a,b]$, as illustrated in Fig. \ref{fig:Resource allocation}. The parameters \(\alpha\) and \(\beta\) define the spacing between active bins in the time and frequency domains, respectively, while \(q_1\) and \(q_2\) specify the starting indices of the active bins in these domains, enabling flexible and efficient resource allocation without degrading OTFS capabilities of representing delay and Doppler efficiently since the same bandwidth and symbol duration is preserved.
 
\subsection{SC-IFDM/OTFS and OFDM Co-existence}

\begin{figure}[t]
    \includegraphics[width=0.4815\textwidth]{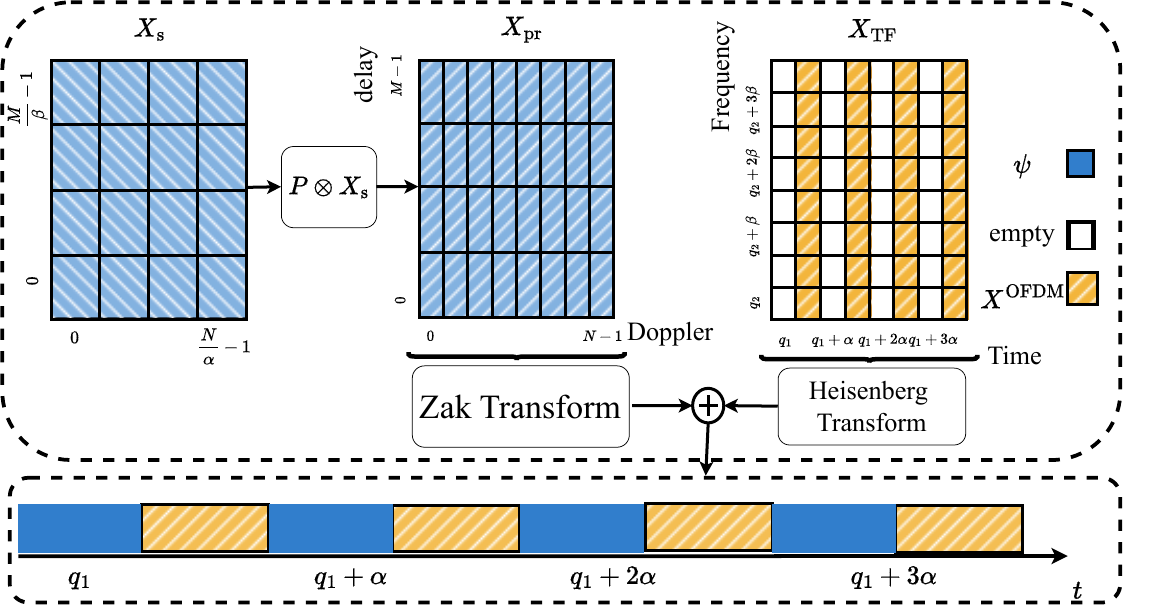}
    \caption{SC-IFDM/OTFS and OFDM co-existence in time ($\alpha=2$). }
\label{fig:OFDM_OTFS_CO}
\end{figure}

\hspace{\parindent}Building on the earlier discussion of waveform flexibility within the \ac{SC-IFDM} framework, this section addresses the coexistence of \ac{SC-IFDM}/\ac{OTFS} with \ac{OFDM} in a multi-user scenario. Given that \ac{OTFS} and \ac{SC-IFDM} are capable of handling high mobility challenges, they are ideal for mobile users. In contrast, \ac{OFDM} remains highly effective for static users due to its low-complexity frequency-domain equalization and reduced system latency. Ensuring the seamless integration of these waveforms within the same system is critical for optimizing the overall performance. To achieve this, an orthogonal co-existence scheme is necessary, allowing \ac{OFDM} to serve static users while \ac{SC-IFDM} or \ac{OTFS} addresses the needs of mobile users by operating efficiently in the \ac{DD} domain. In this context, we focus on \ac{OTFS} as the representative waveform for mobile users.
An orthogonal time co-existence approach, combined with the \ac{FCP} method described in \cite{zegrar2022effect}, is used to address interference between OFDM and OTFS.
In this scheme, \ac{OTFS} occupies $(\frac{MN} {\alpha})$ data bins after setting $\beta=1$ and $q_1=0$, while \ac{OFDM} utilizes the remaining $(MN-\frac{MN}{\alpha})$ data bins. This configuration results in ($N/\alpha$) symbols for \ac{OFDM}, each with an \ac{IFFT} size of $M$, as illustrated in Fig. \ref{fig:OFDM_OTFS_CO}. The \ac{TF} grid for this scheme is populated as follows 
\begin{equation}
\begin{aligned}
X_{\text{TF}}[n,m]=
    \begin{cases}
    0, ~~ &[n]_{\alpha}=0,\\
    X^{\text{OFDM}}(n,m),   & \text{Otherwise}.  
    \end{cases}.
    \end{aligned}
\end{equation}

In this approach, \ac{OFDM} employs \ac{IFFT} to directly convert data into time, as described in \eqref{equ:ofdm}, while \ac{OTFS} is generated through its precoded signal defined in \eqref{equ:otfs_dft} and \eqref{equ:precoded_OTFS} as follows
\begin{equation}
\begin{aligned}
    \mathfrak{s}^{\text{OTFS}}(l+nM)=\frac{1}{\sqrt{N}} \sum_{k=0}^{N-1} X^{\text{OTFS}}_{pr}(k,l) e^{j 2 \pi\frac{k(l+nM)}{MN}} ,
\end{aligned}
\end{equation}
where $X^{\text{OTFS}}_{pr}(k,l)=X^{\text{OTFS}}([k]_{\frac{N}{\alpha}},l)e^{j2\pi(\frac{-a }{\alpha})}e^{j2\pi(\frac{-kl}{MN})}$. After transforming $X_{\text{TF}}$ to time domain using the Heisenberg transform in \eqref{equ:heisenberg} and setting $l=m$, the resulting output signal, $\mathfrak{s}_{\text{OFDM}}^{\text{OTFS}}$, is obtained as
\begin{equation}
\begin{aligned}
\mathfrak{s}_{\text{OFDM}}^{\text{OTFS}}(l+nM)=
    \begin{cases}
    \mathfrak{s}^{\text{OTFS}}(l+nM), ~~ &[n]_{\alpha}=0, \\
    \mathfrak{s}^{\text{OFDM}}(l+nM),   & \text{Otherwise}.  
    \end{cases}.
    \end{aligned}
\end{equation}

The combined signal for both \ac{OTFS} and \ac{OFDM} is then formed, ensuring that each waveform operates in its designated \ac{TF} resources without interference. This method enables efficient co-existence by leveraging the unique strengths of \ac{OTFS} in handling mobility and doubly selective channels, while \ac{OFDM} serves static users with low-complexity processing. After the transmission of the combined signal, the received signal undergoes specialized processing to separate and equalize the two channel.


\subsubsection{OTFS received signal}
The received combined signal, $y_{\text{OFDM}}^{\text{OTFS}}$, after propagating through the doubly selective channel defined in \eqref{equ:Channel_gen}, undergoes initial processing to isolate the \ac{OTFS} contribution. This involves removing the \ac{FCP} and applying a weighting vector to isolate \ac{OTFS} signal. The weighted \ac{OTFS} signal is then expressed as
\begin{equation}
     \breve{y}_{\text{OFDM}}^{\text{OTFS}}(l+nM)=y_{\text{OFDM}}^{\text{OTFS}}(l+nM) . G_1(l+nM) ,
\end{equation}
where $G_1(l+nM)$ is a binary weighting vector defined as 
\begin{equation}
\begin{aligned}
G_1(l+nM)=
    \begin{cases}
    1, ~~ &[n]_{\alpha}=0,\\
    0,   & \text{Otherwise}.  
    \end{cases}
    \end{aligned}
\end{equation}

This operation effectively filters out the \ac{OFDM} components, leaving the \ac{OTFS} signal clean. The \ac{DD} domain \ac{OTFS} signal, $Y^{\text{OTFS}}_{DD}$, can be expressed as  
\begin{equation}
\begin{aligned}
    Y^{\text{OTFS}}_{DD}(k,l)=\sum_{r=0}^{R-1}&h_r e^{j2\pi k_r\frac{(L_{FCP}+l-l_r)}{(M+L_{FCP})N}}\\
    &X^{\text{OTFS}}_{pr}\left([k-k_r]_N,[l-l_r]_M\right),
    \end{aligned}
\end{equation}
where $L_{FCP}$ is the \ac{FCP} length. The received \ac{OTFS} signal is thus free of interference from \ac{OFDM} and can be processed using any known equalization methods. The effective \ac{DD} channel will exhibit $\alpha$ replicas of channel and any standard equalization technique can be applied to recover the transmitted data.

\subsubsection{OFDM received signal}
For the static channel \ac{OFDM} user, after passing through a frequency selective channel (where all $k_r =0 $ in \eqref{equ:Channel_gen}), the received signal, $y_{\text{OFDM}}^{\text{SC-IFDM}}$, is similarly processed. By discarding the \ac{FCP} and applying different weighting vector, the \ac{OFDM} signal can be expressed as 
\begin{equation}
     \breve{y}^{\text{OFDM}}(l+nM)=y_{\text{OFDM}}^{\text{OTFS}}(l+nM) . G_2(l+nM) ,
\end{equation}
where $G_2(l+nM)$ is a weighting vector defined as 
\begin{equation}
\begin{aligned}
G_2(l+nM)=
    \begin{cases}
    0, ~~ &[n]_{\alpha}=0,\\
    1,   & \text{Otherwise}.  
    \end{cases}
    \end{aligned}
\end{equation}

This process effectively removes the \ac{OTFS} components from the signal, isolating the \ac{OFDM} portion. The resulting frequency-domain \ac{OFDM} signal, \( Y_{\text{Freq}}^{\text{OFDM}} \), can be written as
\begin{equation}
\begin{aligned}
    Y_{Freq}^{\text{OFDM}}(n,m)=  
    \begin{cases}
    0, &[n]_{\alpha}=0,\\
    \sum_{r=0}^{R-1}\bar{h_r}X^{\text{OFDM}}(n,m),   & \text{Otherwise},  
    \end{cases}
    \end{aligned}
\end{equation}
where $\bar{h_r}=h_r e^{\frac{-j2\pi m_r n}{M}}$.
This received \ac{OFDM} signal can then be processed using standard equalization techniques such as \ac{MMSE} or \ac{ZF}, ensuring that the signal is free from any \ac{OTFS} interference and can be equalized effectively.

\subsection{SC-IFDM/OTFS and AFDM  Co-existence}
\begin{figure}[t]
    \centering
    \subfigure[]{\includegraphics[width=4cm ]{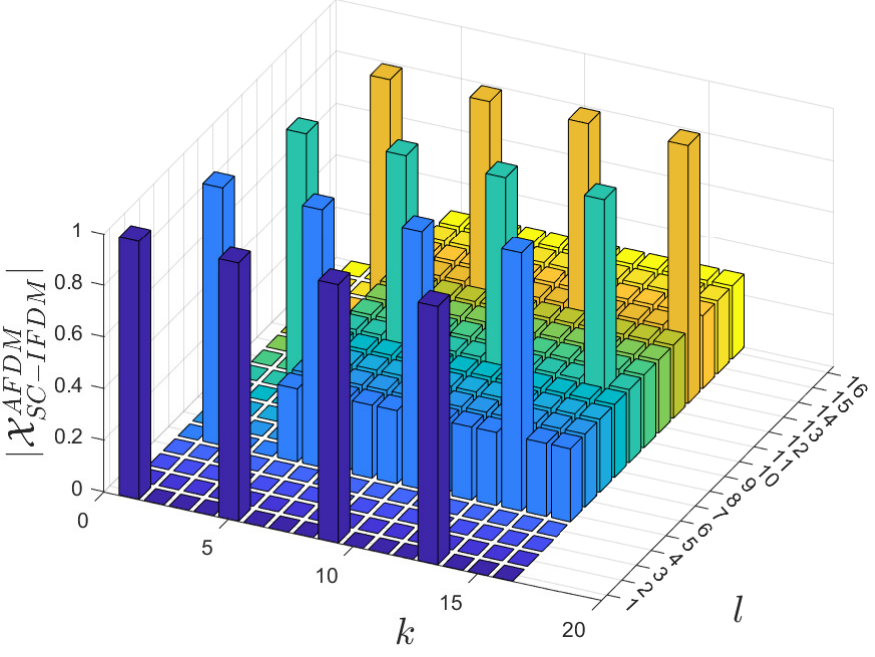}\label{fig:X_SC_AFDM}}
    \subfigure[]{\includegraphics[width=4cm]{X_SC_AFDM.pdf}\label{fig:Y_SC_AFDM}} 
    \caption{ Proposed SC-IFDM-AFDM in SC-IFDM domain ($c_1^{'} = 4, M=16, N=16$), (a) transmitter side (b) receiver side.}
\end{figure}
\begin{figure*}[t]
    \centering
    \subfigure[]{\includegraphics[width=0.325\textwidth]{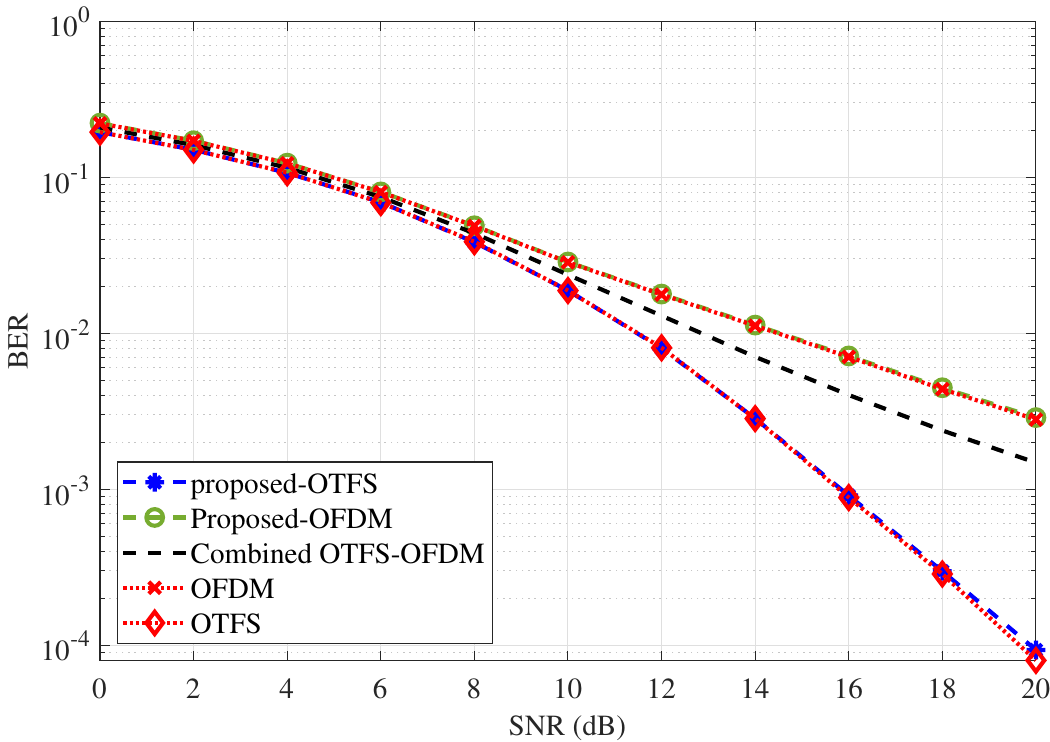}\label{fig:BER_OTFS_OFDM}}
   \subfigure[]{\includegraphics[width=0.325\textwidth]{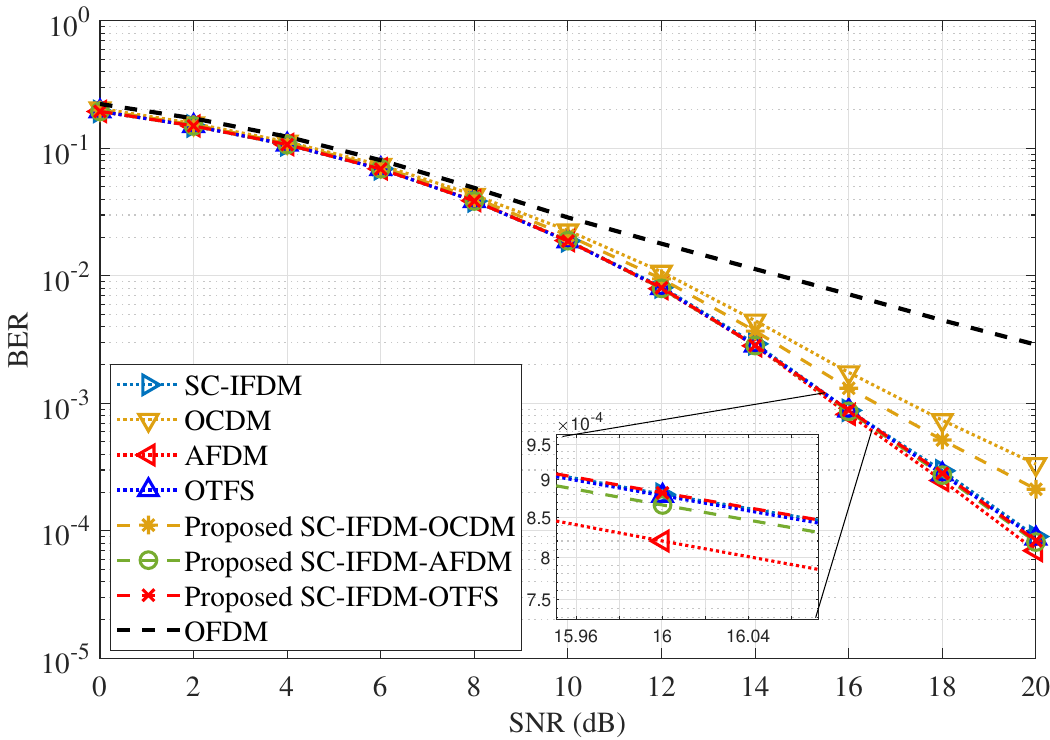}\label{fig:BER}} 
    \subfigure[]{\includegraphics[width=0.325\textwidth]{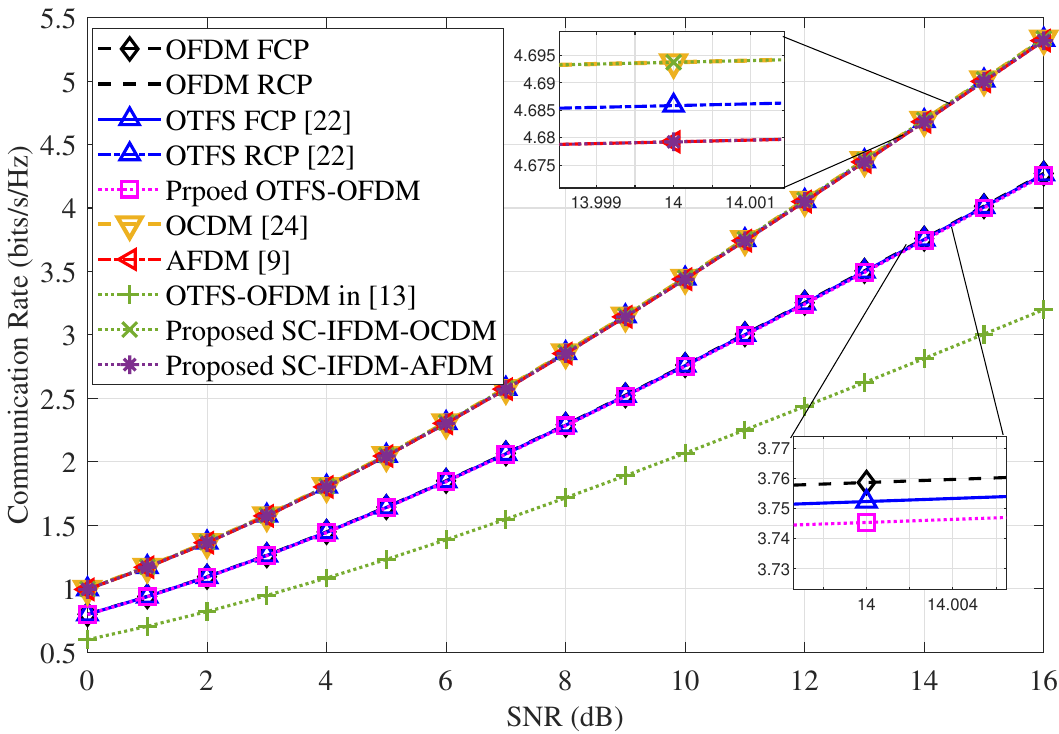}\label{fig:Spectral Efficiency}}
    
    \caption{(a) BER of OTFS/SC-IFDM and OFDM co-existence, (b) BER of the proposed co-existence of SC-IFDM with different waveforms, (c) communication rate vs. SNR analysis.}
    \label{fig:Communication performance}
\end{figure*}
\hspace{\parindent}In this subsection, we explore the co-existence of \ac{SC-IFDM}/\ac{OTFS} and \ac{AFDM} waveforms within a \ac{JSAC} system. In such scenarios, \ac{OTFS}/\ac{SC-IFDM} is employed for communication in dynamic environments, while orthogonal twisted chirps from \ac{AFDM} are used for sensing or communication, depending on the \ac{AFDM} receiver's requirements. The number of twisted chirps can be flexibly allocated based on the needs of both sensing and communication tasks, allowing dynamic resource slicing. Either \ac{OTFS} or \ac{SC-IFDM} can be generated based on the waveform selected by the receiver. For demonstration, we use \ac{SC-IFDM} to illustrate orthogonal co-existence, expressed as
\begin{equation}
\begin{aligned}
    &\mathcal{X}_{\text{SC-IFDM}}^{\text{AFDM}}(k,l)= \\
    &\begin{cases} 
    s^{\text{AFDM},i}(l)e^{-j 2 \pi (\frac{kl}{MN})},&\left[c_1^{'}\frac{-M}{2}-c_1^{'}l+ i-k\right]_N=0, \\ X^{\text{SC-IFDM}}(k,l), & \text {Otherwise}.
    \end{cases}.
\label{equ:SC-IFDM_CO_AFDM}
\end{aligned}
\end{equation}

After transforming $\mathcal{X}_{\text{SC-IFDM}}^{\text{AFDM}}$ to the time domain, a \ac{RCP} is added to the entire frame to avoid discontinuities in \ac{AFDM} signal, as described in \cite{zegrar2022effect}. The received signal at an \ac{SC-IFDM} receiver signal, can be given as 
\begin{equation}
\begin{aligned}
   Y_{\text{SC-IFDM}}^{\text{AFDM}}(k,l)= \sum_{r=0}^{R-1} &h_r e^{j2\pi \frac{k_r}{N}\frac{l-l_r}{M}}\Lambda_r(k,l)\\ &\times \mathcal{X}_{\text{SC-IFDM}}^{\text{AFDM}} ([k-k_r]_N,[l-l_r]_M).
   \end{aligned}
\end{equation}

The twisted chirps, $s^{\text{AFDM},i}$, can also be used for communication channel estimation by using appropriate $c_1$ and leaving a guard in the surrounding of any non-zero point of AFDM as shown in Fig.\ref{fig:X_SC_AFDM} and \ref{fig:Y_SC_AFDM}.
In terms of sensing, the received signal is mixed in time with the conjugate of the allocated $i$-th twisted chirp, $\text{conj}(s^{\text{AFDM},i})$, to estimate the time delay and Doppler shifts introduced by the targets.

It is important to note that \ac{AFDM}, in addition to its use for sensing, can also be employed for communication within this co-existence framework. The dual use of \ac{AFDM} for both sensing and communication allows for efficient resource allocation and performance optimization in dynamic \ac{JSAC} environments.

\section{Simulation Results}\label{sec:simulation}

\hspace{\parindent}In this section, we evaluate the proposed co-existence schemes in terms of communication and sensing for the proposed systems. The \ac{KPIs} used for communication performance are \ac{BER} and \ac{SE}, while for sensing, the KPIs are \ac{RMSE} of the range and velocity. Unless stated otherwise, the key system simulation parameters are summarized in Table~\ref{tab:sensing_simulation_parameters}.

\begin{table}[htbp]
\centering
\caption{Simulation Parameters}
\label{tab:sensing_simulation_parameters}
\begin{tabular}{|l|l|l|}
\hline
\textbf{Description} & \textbf{Parameters} & \textbf{Value} \\ \hline
Carrier frequency & \( f_c \) & 77 GHz \\ \hline
Bandwidth & \( B \) & 200 MHz \\ \hline
Subcarrier spacing & \( \Delta f \) & \( B/M \) MHz \\ \hline
No. of sub-carriers per symbol & \( M \) & 32 \\ \hline
No. of OFDM symbols & \( N \) & 32 \\ \hline
No. of SC-IFDM/OTFS symbols & \( N_{\text{SC-IFDM}} \) & 200  \\ \hline
No. of chirps/SC-IFDM symbol & \( N_c \) & 1\\ \hline
SC-IFDM-AFDM parameter & \(c_{1}^{'}=2\) & 1 \\ \hline
Range resolution & \( \Delta r \) & 0.7495 m \\ \hline
Velocity resolution & \( \Delta v \) & 1.9011 m/s \\ \hline
Range of targets & \( r_r \) & \(\sim U(0, 100)\) m \\ \hline
Velocity of targets & \( v_r \) & \(\sim U(-80, 80)\) m/s \\ \hline
Signal-to-noise ratio & SNR & 15 dB \\ \hline
\end{tabular}
\end{table}
\subsection{Communication Co-existence Performance}

\hspace{1em}First, the \ac{BER} performance of the proposed schemes, including OTFS-OFDM, SC-IFDM-OCDM, SC-IFDM-AFDM, and SC-IFDM-OTFS, is evaluated. The results are illustrated in Fig. \ref{fig:BER_OTFS_OFDM} and Fig. \ref{fig:BER}. The proposed OTFS-OFDM scheme demonstrates a \ac{BER} performance comparable to both the \ac{OTFS} system discussed in \cite{zegrar2022effect} and conventional \ac{OFDM}. This is illustrated in Fig. \ref{fig:BER_OTFS_OFDM}, which depicts both the overall and individual \ac{BER} performance.
A notable feature of this scheme is its integration of \ac{OFDM} within the \ac{OTFS} symbol. This integration enhances the \ac{SE} and reduces latency, allowing \ac{OFDM} users to decode their data without waiting for the entire \ac{OTFS} frame to be processed, particularly when the \ac{OTFS} user requires only ($MN/\alpha$) data.

Fig. \ref{fig:BER} illustrates the \ac{BER} performance of the SC-IFDM-OCDM, SC-IFDM-AFDM, and SC-IFDM-OTFS schemes. SC-IFDM-OCDM demonstrates slightly better \ac{BER} performance compared to conventional \ac{OCDM}, reflecting the inherent advantages of SC-IFDM. Meanwhile, SC-IFDM-AFDM and SC-IFDM-OTFS show \ac{BER} performance comparable to conventional \ac{AFDM} and \ac{OTFS}, respectively.

Next, the (\ac{SE}) of the proposed co-existence waveforms is analyzed. The \ac{SE} depends on parameters such as frame length, cyclic prefix (CP) size, pilot overhead, and guard intervals. Fig. \ref{fig:Spectral Efficiency} illustrates the \ac{SE} of the proposed schemes. The results in Fig. \ref{fig:Spectral Efficiency} show that the proposed OTFS-OFDM scheme achieves higher \ac{SE} than conventional TDD or FDD schemes, as well as the approach proposed in \cite{10382693}, particularly when the OTFS user requires $MN/\alpha$ data. This improvement is attributed to the integration of OFDM within the OTFS symbol, which enables efficient utilization of the \ac{TF} resources without sacrificing delay and Doppler resolution. 
\begin{figure*}[t]
    \centering
    \subfigure[]{\includegraphics[width=0.325\textwidth]{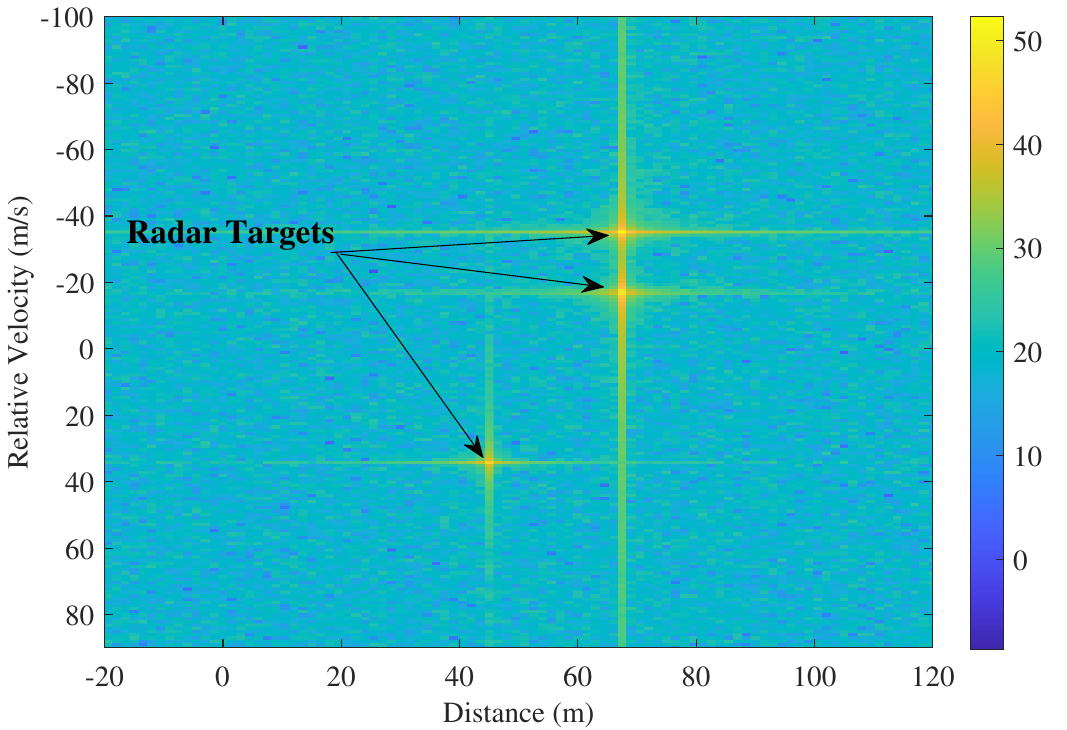}\label{fig:Range_velocity_map}}
   \subfigure[]{\includegraphics[width=0.325\textwidth]{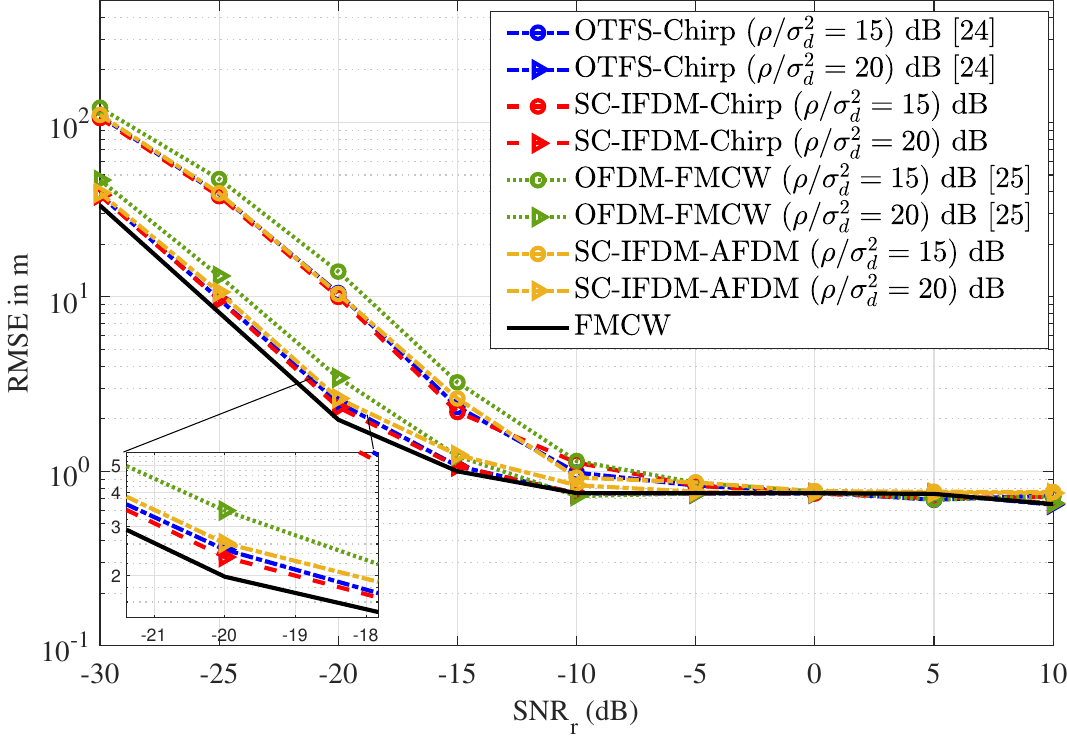}\label{fig:RMSE_R}} 
    \subfigure[]{\includegraphics[width=0.325\textwidth]{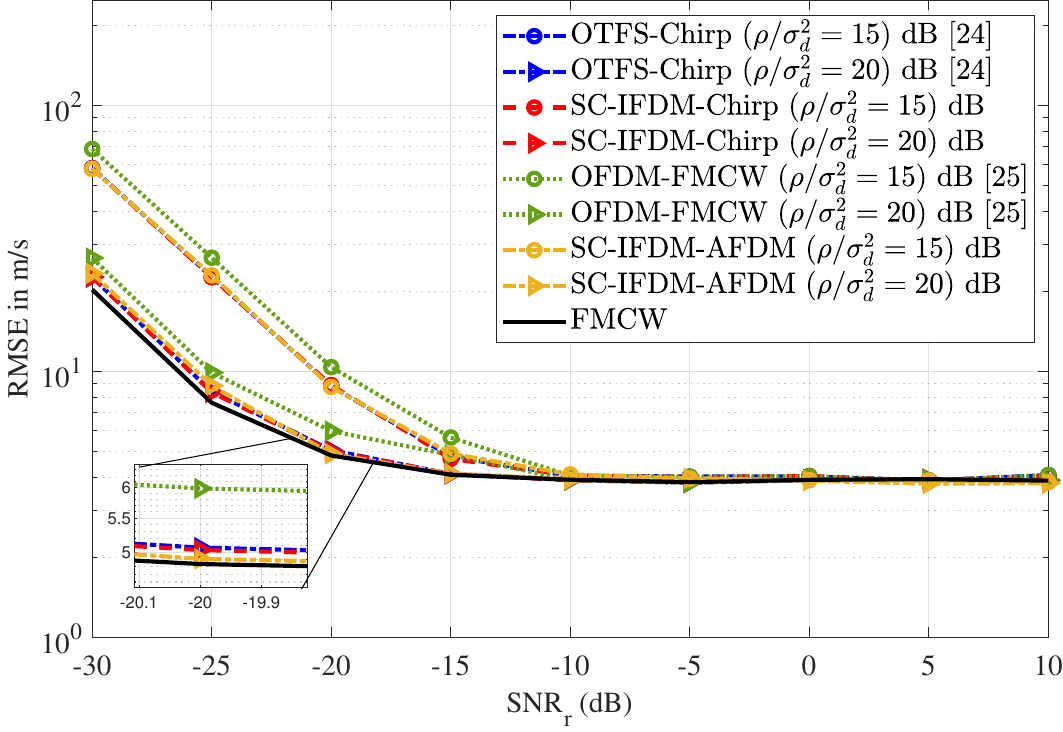}\label{fig:RMSE_V2}}

    \caption{(a) Range-Doppler map (3 targets), (b) range RMSE performance vs SNR (3 targets), (c) velocity RMSE performance vs SNR (3 targets).}
    \label{fig:sens_perf}
\end{figure*}

The \ac{SE} for the OTFS-OFDM scheme is derived considering $MN/\alpha$ data, with the \ac{FCP} ensuring a clean signal. The communication rate for OTFS-OFDM is expressed as
\begin{equation}
\begin{aligned}
\mathcal{C}_\mathrm{OFDM}^{\mathrm{OTFS}} = &\left(\frac{MN/\alpha - \varphi_\mathrm{OTFS}}{MN + NL_\mathrm{FCP}}\right) \log_2 \left( 1 + \gamma \right) \\ 
&+ \left( \frac{T/\alpha - \varphi_\mathrm{OFDM}}{T + T_\mathrm{cp}} \log_2 \left( 1 + \gamma \right) \right),
\end{aligned}
\end{equation}
where \(\gamma\), \(\varphi_\mathrm{OTFS}\), \(\varphi_\mathrm{OFDM}\), and \(\alpha\) denote the communication \ac{SNR}, pilot and guard overhead of \ac{OTFS}, pilot and guard overhead of \ac{OFDM}, and the co-existence ratio, respectively.

Using the \ac{SE} formulas from \cite{zegrar2022effect} and \cite{haif2024novel}, the \ac{SE} for conventional and other proposed waveforms is calculated as follows
\begin{equation}
\mathcal{C}_\mathrm{OFDM} = \left( \frac{T - \vartheta_\mathrm{OFDM}}{T + T_\mathrm{cp}} \log_2 \left( 1 + \gamma \right) \right),
\end{equation}
\begin{equation}
\mathcal{C}_\mathrm{OTFS_{FCP}} = \left( \frac{M - \vartheta_\mathrm{OTFS}}{M + L_\mathrm{FCP}} \log_2 \left( 1 + \gamma \right) \right),
\end{equation}
\begin{equation}
\mathcal{C}_\mathrm{OCDM} = \left( \frac{MN - \vartheta_\mathrm{OCDM}}{MN} \log_2 \left( 1 + \gamma \right) \right),
\end{equation}
\begin{equation}
\mathcal{C}_\mathrm{AFDM} = \left( \frac{MN - \vartheta_\mathrm{AFDM}}{MN} \log_2 \left( 1 + \gamma \right) \right),
\end{equation}
where \(\vartheta_\mathrm{OCDM}\) and \(\vartheta_\mathrm{AFDM}\) represent the pilot and guard overheads specific to \ac{OCDM} and \ac{AFDM}\cite{haif2024novel}, respectively.
Finally, the \ac{SE} of the proposed SC-IFDM-OCDM, SC-IFDM-AFDM, and SC-IFDM-OTFS schemes is analyzed. The \ac{SE} of these waveforms depends on the number of data symbols and pilots. Due to the orthogonality between these waveforms in the proposed co-existence schemes, their \ac{SE} can be directly calculated. As illustrated in Fig. \ref{fig:Spectral Efficiency}, the proposed SC-IFDM-OCDM, SC-IFDM-AFDM, and SC-IFDM-OTFS waveforms demonstrate no loss in \ac{SE} compared to their individual use. This result highlights the effectiveness of the proposed mother waveform in maintaining the \ac{SE} of the constituent waveforms while enabling their coexistence within the same \ac{TF} resources.
\subsection{Sensing Co-Existence Performance}

\hspace{1em}The sensing performance of the proposed waveforms is evaluated using a range and velocity estimation scenario involving three targets. The results are presented in Fig. \ref{fig:sens_perf}, which includes: (a) the range-Doppler map, (b) the range RMSE performance versus \ac{SNR}, and (c) the velocity RMSE performance versus \ac{SNR}. To quantify sensing accuracy, the RMSE of the range and velocity is analyzed for the proposed SC-IFDM-Chirp and SC-IFDM-AFDM waveforms. For comparison, the performance of existing methods, including OTFS-Chirp \cite{zegrar2024novel}, OFDM-FMCW \cite{bouziane2024novel}, and FMCW alone, is also evaluated. For OCDM, which employs multiple orthogonal chirps, the performance of a single chirp is presented to simplify the comparison.

The results, as shown in Fig.~\ref{fig:RMSE_R} and Fig.~\ref{fig:RMSE_V2}, demonstrate that the proposed \ac{SC-IFDM}-Chirp and \ac{SC-IFDM}-AFDM waveforms achieve performance comparable to OTFS-FMCW. This is due to the inherent similarity between \ac{SC-IFDM} and \ac{OTFS} in their ability to represent chirps sparsely, as previously discussed. Furthermore, the proposed schemes exhibit performance close to \ac{FMCW} with appropriate power allocation. The sensing performance of SC-IFDM-Chirp and SC-IFDM-AFDM depends on the power ratio assigned to the chirp signal $\rho/\sigma_{d}^2$, where $\sigma_{d}^2$ is the data power. Simulations conducted with two power ratios ($15\,\mathrm{dB}$ and $20\,\mathrm{dB}$) show that increasing the chirp power ratio significantly improves performance, bringing it closer to that of standalone \ac{FMCW}. This enhancement results from the higher radar \ac{SNR} ($\text{SNR}_r$) of the chirp, which directly improves the accuracy of range and velocity estimation.

In contrast, the existing OFDM-FMCW method exhibits inferior performance. This degradation is primarily due to the loss of orthogonality between the chirp signal and data, as well as among the data symbols themselves. The issue arises because the chirp is extracted from the diagonal elements of the DFT matrix, leading to interference. This interference limits sensing accuracy and highlights the importance of designing better waveforms for such systems.

The results demonstrate that the proposed SC-IFDM-Chirp, SC-IFDM-OCDM, and SC-IFDM-AFDM waveforms provide robust sensing performance comparable to OTFS-FMCW and FMCW, while addressing the limitations of existing methods such as OFDM-FMCW. The flexibility in power allocation and orthogonality preservation highlights their effectiveness for joint communication and sensing applications.


\section{conclusion} \label{sec:conclusion}

\hspace{\parindent}In this paper, we proposed a unified "mother waveform" framework, capable of generating multiple \ac{DFT}-based waveforms like OTFS, OFDM, FMCW, OCDM, and AFDM from a single SC-IFDM structure. We showed that these waveforms differ primarily in phase shifts and data allocation strategies, enabling efficient orthogonal coexistence within shared time-frequency resources. This solution supports diverse 4G, 5G, and 6G applications such as URLLC, mMTC, and high-mobility scenarios, while ensuring backward compatibility with existing infrastructure. Our simulations confirmed the framework's effectiveness in achieving low BER, efficient resource allocation, and robust multi-user performance. Future work may focus on real-time implementation on SDR platforms and using \ac{AI} / \ac{ML}for dynamic waveform switching, advancing the adaptability and scalability of communication systems for 6G networks and beyond.



\end{document}

%% file: acro.tex
\DeclareAcronym{URLLC}{short = URLLC,long  = ultra-reliable low latency communications ,tag    = abbrev}
\DeclareAcronym{mMTC}{short = mMTC,long  = massive machine-type communications ,tag    = abbrev}
\DeclareAcronym{INI}{short = INI,long  = inter-numerology interference ,tag    = abbrev}
\DeclareAcronym{PHY}{short = PHY,long  = physical layer ,tag    = abbrev}
\DeclareAcronym{IoT}{short = IoT,long  = internet of things ,tag    = abbrev}
\DeclareAcronym{eMBB}{short = eMBB,long  = enhanced mobile broadband ,tag    = abbrev}
\DeclareAcronym{MIMO}{short = MIMO,long  = multiple-input multiple-output ,tag    = abbrev}
\DeclareAcronym{ICI}{short = ICI,long  = inter-carrier interference ,tag    = abbrev}
\DeclareAcronym{6G}{short = 6G,long  = 6th generation ,tag    = abbrev}
\DeclareAcronym{5G}{short = 5G,long  = 5th generation ,tag    = abbrev}
\DeclareAcronym{4G}{short = 4G,long  = 4th generation ,tag    = abbrev}
\DeclareAcronym{V2X}{short = V2X,long  = vehicle-to-everything ,tag    = abbrev}
\DeclareAcronym{PAPR}{short = PAPR,long  = peak-to-average power ratio ,tag    = abbrev}
\DeclareAcronym{DFT-S-OFDM}{short = DFT-S-OFDM,long  = \ac{DFT} spread \ac{OFDM} ,tag    = abbrev}
\DeclareAcronym{UW-DFT-S-OFDM}{short = UW-DFT-S-OFDM,long  = unique word \ac{DFT-S-OFDM} ,tag    = abbrev}
\DeclareAcronym{ZT-DFT-S-OFDM}{short = ZT-DFT-S-OFDM,long  = zero-tail \ac{DFT-S-OFDM} ,tag    = abbrev}
\DeclareAcronym{DD}{short = DD,long  = delay-Doppler ,tag    = abbrev}
\DeclareAcronym{ISI}{short = ISI,long  = inter-symbol interference ,tag    = abbrev}
\DeclareAcronym{DFT}{short = DFT,long  = discrete Fourier transform ,tag    = abbrev}
\DeclareAcronym{IDFT}{short = IDFT,long  = inverse discrete Fourier transform ,tag    = abbrev}
\DeclareAcronym{IFFT}{short = IFFT,long  = inverse fast Fourier transform ,tag    = abbrev}
\DeclareAcronym{AWGN}{short = AWGN,long  = additive white Gaussian noise,tag    = abbrev}
\DeclareAcronym{FMCW}{short = FMCW,long  = frequency modulated continuous wave ,tag    = abbrev}
\DeclareAcronym{OFDM}{short = OFDM,long  = orthogonal frequency division multiplexing ,tag    = abbrev}
\DeclareAcronym{UW-OFDM}{short = UW-OFDM,long  = unique word \ac{OFDM} ,tag    = abbrev}
\DeclareAcronym{ZT-OFDM}{short = ZT-OFDM,long  = zero-tail \ac{OFDM} ,tag    = abbrev}
\DeclareAcronym{GFDM}{short = GFDM,long  = generalized frequency division multiplexing ,tag    = abbrev}
\DeclareAcronym{SC-IFDM}{short = SC-IFDM,long  = single-carrier interleaved frequency division multiplexing ,tag    = abbrev}
\DeclareAcronym{OCDM}{  short = OCDM,  long  = orthogonal chirp division multiplexing ,  tag    = abbrev}
\DeclareAcronym{OC}{  short = OC,  long  = orthogonal chirp ,  tag    = abbrev}
\DeclareAcronym{DZT}{  short = DZT,  long  = discrete Zak transform ,  tag    = abbrev}
\DeclareAcronym{AFDM}{  short = AFDM,  long  = affine frequency division multiplexing,  tag    = abbrev}
\DeclareAcronym{OTFS}{  short = OTFS,  long  = orthogonal time-frequency space ,  tag    = abbrev}
\DeclareAcronym{IDZT}{  short = IDZT,  long  = inverse discrete Zak transform ,  tag    = abbrev}
\DeclareAcronym{ISFFT}{  short = ISFFT,  long  = inverse symplectic finite Fourier transform,  tag    = abbrev}
\DeclareAcronym{SFFT}{  short = SFFT,  long  = symplectic finite Fourier transform ,  tag    = abbrev}
\DeclareAcronym{IDFnT}{  short = IDFnT,  long  = inverse discrete Fresnel transform ,  tag    = abbrev}
\DeclareAcronym{DFnT}{  short = DFnT,  long  = discrete Fresnel transform ,  tag    = abbrev}
\DeclareAcronym{IDAFT}{  short = IDAFT,  long  = inverse discrete affine Fourier transform,  tag    = abbrev}
\DeclareAcronym{DAFT}{  short = DAFT,  long  = discrete affine Fourier transform,  tag    = abbrev}
\DeclareAcronym{JSAC}{  short = JSAC,  long  = joint sensing and communication   ,  tag    = abbrev}
\DeclareAcronym{CP}{  short = CP,  long  = cyclic prefix  ,  tag    = abbrev}
\DeclareAcronym{RCP}{  short = RCP,  long  = reduced cyclic prefix   ,  tag    = abbrev}
\DeclareAcronym{FCP}{  short = FCP,  long  = full cyclic prefix   ,  tag    = abbrev}
\DeclareAcronym{BER}{  short = BER,  long  = bit error rate ,  tag    = abbrev}
\DeclareAcronym{RMSE}{  short = RMSE,  long  = root mean squared error ,  tag    = abbrev}
\DeclareAcronym{TF}{  short = TF,  long  = time-frequency ,  tag    = abbrev}
\DeclareAcronym{ZF}{  short = ZF,  long  = zero forcing ,  tag    = abbrev}
\DeclareAcronym{MMSE}{  short = MMSE,  long  = minimum mean square error ,  tag    = abbrev}

\DeclareAcronym{SNR}{  short = SNR,  long  = signal-to-noise ratio ,  tag    = abbrev}

\DeclareAcronym{SE}{  short = SE,  long  = spectral efficiency ,  tag    = abbrev}

\DeclareAcronym{KPIs}{  short = KPIs,  long  = key performance indicators ,  tag    = abbrev}

\DeclareAcronym{AI}{  short = AI,  long  = artificial intelligence ,  tag    = abbrev}

\DeclareAcronym{ML}{  short =  ML,  long  = machine learning,  tag    = abbrev}